\documentclass[12pt,preprint]{aastex}

\newcommand\Ha{\mathrm{H}\alpha}
\newcommand\Hb{\mathrm{H}\beta}

\newcommand\co{$^{12}$CO}
\newcommand\Ja{(\textit{J}\ =\ 1--0)}

\slugcomment{Not to appear in Nonlearned J., 45.}

\shorttitle{CO(J=1--0) Survey of Nearby Galactic Centers I}
\shortauthors{Komugi et al.}

\begin{document}


\title{Molecular Gas
Distribution in Barred and Non-barred Galaxies along the Hubble Sequence}


\author{S. Komugi\altaffilmark{1,2,3}, Y.Sofue\altaffilmark{1,4}, K.Kohno\altaffilmark{1},
H.Nakanishi\altaffilmark{4,5}, S.Onodera\altaffilmark{1},
F.Egusa\altaffilmark{1,3},
K.Muraoka\altaffilmark{1,3}}


\altaffiltext{1}{Institute of Astronomy, University of Tokyo}
\altaffiltext{2}{National Astronomical Observatory of Japan}
\altaffiltext{3}{JSPS Research Fellowship for Young Scientists}
\altaffiltext{4}{Kagoshima University}
\altaffiltext{5}{Australia Telescope National Facility / CSIRO}


\begin{abstract}
We present results from a survey of $^{12}$CO\ \Ja \ \ spectra obtained for the central regions of 68 nearby
galaxies at an angular resolution of 16$^{\prime \prime}$ using the Nobeyama Radio Observatory 45m telescope, aimed at
 characterizing the properties of star forming molecular gas.  Combined with similar
resolution observations in the literature, the compiled sample set of 166
galaxies span a wide range of galactic properties.
NGC\ 4380, which was previously undetected in CO, was detected.

This initial paper of a series will focus on the data and the gaseous properties of the samples, and
 particularly on the degree of central concentration of molecular gas in
 a range of morphological types, from early (S0/Sa) to late (Sd/Sm)
 galaxies with and without bars. 
The degree of molecular central concentration
 in the central kiloparsec, compared to the central several kiloparsecs of galaxies, is found to vary smoothly with Hubble
 type, so that early type galaxies show larger central concentration.  The comparison of barred and non-barred galaxies within 
early and late type galaxies suggest that difference in Hubble type,
 representing the effect of bulges, is the more important factor in concentrating gas into the central regions than bars.

\end{abstract}


\section{Introduction}

Molecular gas is the driving element of star formation in galaxies, and
its quantitative relation to the current star formation rate is
essential to our understanding of galactic evolution.  It has become
increasingly clear that the dynamical properties of galaxies
are responsible for redistributing molecular gas on large scales, and hence the star
formation which follows is also affected by the dynamical properties.
These dynamical properties are expressed as large scale features
(morphology) of the galaxies, i.e., the presence bars and difference
between Hubble type.
These two features (presence of bars and Hubble type) are most prominent and characteristic in the central kiloparsecs of galaxies,
which is also where most of the molecular gas in galaxies is present.
Therefore, the distribution of gas within the central region of various morphologies and its
 relation with star formation has gained
attention in recent years.  

Bars on kpc scales at the centers of galaxies are known to redistribute
angular momentum and transfer molecular gas to the centers.
\cite{sakamoto99b} showed that molecular gas in barred galaxies has a higher degree of concentration in its 
central 1 kpc compared to gas inside the optical disk, by using 
interferometric data from the Nobeyama Millimeter Array (NMA) and Owens Valley Radio Observatory (OVRO) at $4^{\prime \prime}$ resolution.  
Similarly, \cite{sheth}
showed by using the Berkeley-Illinois-Maryland-Array (BIMA) at $6^{\prime \prime}$ resolution that the central 1 kpc of 
barred galaxies has four times more molecular gas compared to the global average.  Some barred galaxies in their sample were not
detected, and they attribute it to the quenching of gas due to efficient star formation in barred galaxies.
Single dish observations have also proved effective.  By using the Nobeyama Radio Observatory$^*$ (NRO) 45m
 telescope, \cite{kuno} mapped out 40 nearby galaxies at $16^{\prime \prime}$ resolution.
They defined the "inner" central concentration $f_\mathrm{in}$ as the ratio between molecular mass in $1/8$ of the
 K band radius (defined at 20 $\mathrm{mag\ arcsec^{-1}}$ in the $K_S$ band) and that within $1/2$ of the K band radius.  They similarly defined the "outer" central 
concentration $f_\mathrm{out}$ as the ratio between molecular mass in $1/2$ of the K band radius and that within the whole of the 
K band radius.  They found that barred galaxies have a significantly higher $f_\mathrm{in}$, whereas the $f_\mathrm{out}$ did not
show much dependence.  Furthermore, they found a correlation between $f_\mathrm{in}$ and the bar strength.

It seems that molecular gas concentration due to bars is a well established phenomenon, and this has also been explored
in terms of numerical simulations(\cite{combes85}; \cite{pfenniger}; \cite{athanassoula}; \cite{wada}).

  Molecular gas is also known to be distributed differently
between early and late type galaxies.  
Molecular gas mass in the
centers of galaxies have been found to decrease in later Hubble types (e.g.,
\cite{young89}, \citet{boeker}).  This has been attributed
to the existence of a stellar bulge in early type galaxies (\cite{young89}) creating a deep potential well in which
gas is accumulated.  Even in the absence of stellar bulges, possible correlations with nuclear star
clusters and stellar surface brightness have been found (\cite{boeker}).  Sheth et al. (2005) report a four-fold central molecular 
mass increase in barred early type galaxies compared to barred late type galaxies.  Early type galaxies, however, have been found to
host stronger bars than late types, with larger $m=2$ and $m=4$ fourier amplitudes (\cite{laurikainen}, but see also
\cite{laurikainen04}).  This points to the possibility that early type galaxies may show larger central concentration simply
because they host stronger bars.
It is important, therefore, to quantify which factor (the presence of bars or Hubble
type) plays the larger role in ridistributing gas to the central regions of galaxies.
  
Mapping observations in CO are optimum for the study of molecular gas
distribution in nearby galaxies, but available galaxy samples are limited by the vast observing
time needed for such projects (e.g., Sakamoto et al. 1999; Helfer et al. 2001;
Sofue et al. 2003).  In turn, single beam measurements at the
centers of a large sample of galaxies lack the precision in terms of
spatial distribution of molecular gas, but can compensate for the lack of
number of sample galaxies.  In this respect, quantifying the CO content of nearby galactic
centers for a large sample using single dish telescopes, is still rewarding.

In order to complement existing studies on molecular gas concentration in galaxies of various morphology,
we have conducted a large survey of the central region of nearby galaxies using 
NRO 45m telescope.

Our data provides a unique opportunity to study the gas
distribution in the central region, with several advantages:  1) Samples are
distributed mostly at the distance of 16Mpc, and therefore the single
beam subtends a similar scale of the central 1 kpc of galaxies.  2) Many
of our samples are also observed by Young et al. (1995) using the FCRAO
14m dish, with an angular resolution of 45$^{\prime \prime}$ or a linear
scale of 3 kpc.  Therefore, the ratio of CO intensity measured by the different surveys
can be interpreted as a degree of central concentration of molecular gas.  

\section{Sample Selection}

The sample galaxies have been selected from those observed and detected in the
FCRAO catalog (\citet{young95} and \citet{KY88}) of nearby galaxies.
The FCRAO survey is the largest CO
 survey of nearby galaxies (300 galaxies), 
 spanning a wide range of parameter space for spirals, but
with a large beamsize of 45$^{\prime \prime}$.  
All of our observing samples, mainly spirals, come from galaxies observed in this survey.    
Since one of the secondary aims of our survey is compare the CO data with star formation (to be presented in 
forthcoming papers), galaxies in the FCRAO survey were selected on whether they had available $\Ha$ data as
 listed in \cite{young96}, and also with measured $\Ha /\Hb$ ratio (to be used to correct for internal dust 
extinction) in \citet{Ho97d}.  Ninety-one galaxies meet this criteria, of which 14 were already
observed at either the NRO 45m or the IRAM 30m telescope which has a similar beamsize as the NRO 45m
telescope.  Therefore, the final selected sample consisted of 77 galaxies.  Galaxies in the FCRAO survey were selected to span
a wide variety of galaxies, and were not magnitude or volume limited.  Our sample also follows this idea.

In the observation (see following section), 38 of the 77 were actually observed, due mainly to weather conditions.
An additional five, with $\Ha$ data from \cite{young96} but without $\Ha /\Hb$, and 25 with observations from the
FCRAO survey but not $\Ha$, were observed.  These galaxies without $\Ha$ measurements were simply selected by whether they
were observable at that certain perioud.  Thus, our observed sample consists of 68 galaxies.
Table \ref{45msample} lists the observed samples.

\footnotetext{Nobeyama Radio Observatory is a branch of the National
Astronomical Observatory of Japan,  National Institutes of Natural Sciences.}

\clearpage

\begin{deluxetable}{llll}
\footnotesize
\tablecolumns{4}
\tablecaption{Galaxies observed at the NRO45m telescope. \label{45msample}}
\tablewidth{0pt}
\tablehead{
\colhead{Galaxy} & \colhead{R.A.}   & \colhead{DEC}   & \colhead{Distance} \\
\colhead{}       & \colhead{J2000}  & \colhead{J2000} & \colhead{Mpc} \\
\colhead{(1)}    & \colhead{(2)}    & \colhead{(3)}   & \colhead{(4)}  } 
\startdata 
 NGC\ 253  & 00 47 35.2	       & -25 17 20 	        &       2.2\\
 NGC\ 520  & 01 24 34.7        & +03 47 49.8         &       30.3\\
 NGC\ 772  & 01 59 20.3	       & +19 00 22 	&	34.1\\
 NGC\ 877  & 02 17 58.7        & +14 32 50.3 	&	54.9 \\
 NGC\ 908  & 02 23 04.8$^{**}$    & -21 14 06 	&	19.6 \\
 NGC\ 1022 & 02 38 32.9$^{**}$    & -06 40 29 	&	20.1 \\
 NGC\ 1482 & 03 54 39.4$^{**}$    & -20 30 08.5 	        &	20.5 \\
 NGC\ 1569 & 04 30 50.3	      & +64 50 48         & 	3.1  \\
 NGC\ 1961 & 05 42 04.3	      & +69 22 46 	        &	54.1 \\
 NGC\ 2276 & 07 27 14.4$^*$    & +85 45 16		&	34.4 \\
 NGC\ 2336 & 07 27 04.1$^*$    & +80 10 41 	&       31.9 \\
 NGC\ 2339 & 07 08 20.5$^*$    & +18 46 49 	&       31.1 \\
 NGC\ 2403 & 07 36 54.5       & +65 35 58.4        &	2.2 \\
 NGC\ 2559 & 08 17 07.5$^{**}$ & -27 27 34 	&       17.3 \\
 NGC\ 2683 & 08 52 41.3$^*$    & +33 25 19 	&       3.2 \\
 NGC\ 2976 & 09 47 15.6        & +67 54 49 	&       2.3 \\
 NGC\ 3034 & 09 55 54.0	      & +69 40 57 	&       2.2 \\
 NGC\ 3147 & 10 16 53.6$^*$    & +73 24 03 	&       38.4 \\
 NGC\ 3556 & 11 11 30.9    & +55 40 27 	        &	10.3 \\
 NGC\ 3593 & 11 14 37.0$^*$    & +12 49 04 	&       7.3 \\
 NGC\ 3623 & 11 18 56.0$^*$    & +13 05 32.0 	&       4.5 \\
 NGC\ 3631 & 11 21 02.9$^*$    & +53 10 11            &	16.6 \\
 NGC\ 3675 & 11 26 08.6$^*$    & +43 35 09 	&	9.8 \\
 NGC\ 3690 & 11 28 33.6        & +58 33 51.3           &       41.4 \\
 NGC\ 3893 & 11 48 39.1        & +48 42 40 	&	13.8 \\
 NGC\ 3938 & 11 52 49.8 	      & +44 07 26       &       11.2 \\
 NGC\ 4038 & 12 01 53.0$^{**}$ & -18 51 48       &       19.2 \\
 NGC\ 4039 & 12 01 54.2$^{**}$ & -18 53 06       &       19.2 \\
 NGC\ 4041 & 12 02 12.2$^{*}$  & +62 08 14	&	17.6    \\
 NGC\ 4088 & 12 05 35.3        & +50 32 31 	&	16.4 \\
 NGC\ 4096 & 12 06 01.0        & +47 28 31 	&       11.2 \\
 NGC\ 4157 & 12 11 04.4$^*$    & +50 29 05 &       12.1  \\
 NGC\ 4178 & 12 12 46.4$^*$    & +10 51 57  &	16.1 \\
 NGC\ 4212 & 12 15 39.3$^*$    & +13 54 06	&	16.1 \\
 NGC\ 4293 & 12 21 12.9$^*$ & +18 22 57	&	16.1 \\
 NGC\ 4298 & 12 21 32.8$^*$    & +14 36 22		&	16.1 \\ 
 NGC\ 4302 & 12 21 42.4        & +14 36 05 		&	16.1 \\
 NGC\ 4321 & 12 22 54.9$^*$    & +15 49 20 		&	16.1 \\
 NGC\ 4380 & 12 25 22.2$^*$    & +10 01 00      	&       16.1 \\
 NGC\ 4394 & 12 25 55.6$^*$    & +18 12 50 	&	16.1 \\
 NGC\ 4418 & 12 26 54.2        & -00 52 45        	&	25.5 \\
 NGC\ 4419 & 12 26 56.4$^*$    & +15 02 51		&	16.1 \\\
 NGC\ 4424 & 12 27 11.6$^*$    & +09 25 14      &      16.1 \\
%% NGC\ 4459 & 12 29 00.0$^*$    & +13 58 43 	&	16.1 \\
 NGC\ 4527 & 12 34 08.8        & +02 39 13         &	16.1 \\
 NGC\ 4536 & 12 34 27.1$^*$    & +02 11 16 	&	16.1 \\
 NGC\ 4548 & 12 35 55.2$^*$    & +14 29 47 	&       16.1 \\
 NGC\ 4567 & 12 36 32.7$^*$    & +11 15 29 		&	16.1 \\ 
NGC\ 4568 & 12 36 34.7        & +11 14 15 	&	16.1 \\
 NGC\ 4647 & 12 43 32.3$^*$    & +11 34 55 	&	16.1 \\
 NGC\ 4666 & 12 45 08.9        & -00 27 38 	&	18.6 \\
 NGC\ 4689 & 12 47 45.5$^*$    & +13 45 46 	&       16.1 \\
 NGC\ 4691 & 12 48 14.1$^{**}$ & -03 19 51 	&	13.1 \\
 NGC\ 4698 & 12 48 23.5        & +08 29 16	&	16.1 \\
 NGC\ 4710 & 12 49 38.9$^*$    & +15 09 56	        &	16.1 \\
 NGC\ 4818 & 12 56 48.7$^{**}$ & -08 31 25 	&	13.5 \\
 NGC\ 4845 & 12 58 01.4        & +01 34 30     	&       16.1 \\
 NGC\ 4984 & 13 08 57.3$^{**}$ & -15 30 59 	&	14.7 \\
 NGC\ 5236 & 13 36 59.4        & -29 52 04 	&	5.9 \\
 NGC\ 5247 & 13 38 03.5        & -17 53 02 	&       20.1 \\
 NGC\ 5713 & 14 40 11.5        & -00 17 26 &       24.6 \\
 NGC\ 5861 & 15 09 16.5$^{**}$ & -11 19 24 		&	24.1 \\
 NGC\ 6000 & 15 49 49.4 	      & -29 23 13 	&	27.1 \\
 NGC\ 6814 & 19 42 40.1$^{**}$ & -10 19 27 	&	21.1 \\
 NGC\ 6946 & 20 34 51.9 	      & +60 09 15 	&	6.7 \\
 NGC\ 6951 & 20 37 15.2 	      & +66 06 22 &	21.7 \\
 NGC\ 7479 & 23 04 57.1        & +12 19 18 	&	34.7 \\
 NGC\ 7541 & 23 14 43.0        & +04 32 05 &      38.1 \\
 IC\  342  & 03 46 49.7        & +68 05 45 &       3.0  \\ \hline \hline
%% IC\  694  & 11 28 31.3 	      & +58 33 28  &	41.4 \\ \hline \hline
\enddata
\tablenotetext{(1)}{Galaxy name.} 
\tablenotetext{(2)(3)}{Right ascension and declination in J2000.0 epoch, from Dressel 
 \& Condon (1976) unless otherwise noted.  Coordinates
with $*$ from NED, $**$ from RC2. 
 All of the  coordinates of the different references were found to be consistent
within the beam size of 
the FCRAO survey, 
45$^{\prime \prime}$, and mostly within 6$^{\prime \prime}$, 
except for NGC\ 1022, NGC\ 2559, and NGC\ 4038.}
%\tablenotetext{(4)}{Morphological type, as listed in \citet{young95}, mainly from RC2.}
\tablenotetext{(4)}{Distance as listed in \citet{young95}, scaled using
 Hubble constant
 $H_o=75\ \mathrm{km\ s^{-1}\ Mpc^{-1}}$.  NGC\ 4874 from
 \citet{liu}.  For
 members of the Virgo cluster, 16.1 Mpc is assumed from Cepheid
 calibrations (\cite{ferrarese}).} 
\end{deluxetable}

\clearpage

\section{Observations}
The observations of the \co\ \Ja \ line (rest frequency 115.27 GHz) were conducted with the 45m telescope at the Nobeyama Radio 
Observatory (NRO) of the National Astronomical Observatory of Japan,
 during three observing runs in May 2005,  
December 2005, and January 2006, under moderate weather conditions.  Two Superconductor-Insulator-Superconductor (SIS)
 receivers, S100 and/or S80, were used for the observations. 
 The backends were the 1024 channel 
digital auto-correlator with a frequency coverage of 512 MHz, or the
 2048 channel acousto-optical spectrometer with a frequency coverage of
 250 MHz.  The CO
line was observed in the upper sideband.
  Typical system temperatures ranged from 700 to 2000 K in the May 2005
 run, and 450 to 1000 K (all in single-sideband scale) in the December 2005 and January 2006 run.
The beamwidth was 16$^{\prime \prime}$.  Pointing was checked using SiO masers or
 continuum sources at 43 GHz every 0.5--1 hour, and found to be accurate within 5$^{\prime \prime}$ unless 
otherwise noted where wind prevented pointing at such accuracy.  Each of the target galaxies was observed with 
a single beam at the center.  

The obtained data  were reduced with the NEWSTAR software used commonly at NRO, which is based on the Astronomical Image
Processing System (AIPS) package, developed by the National Radio Astronomy Observatory. 
 After flagging bad spectra, first or second order baselines were subtracted, and then the antenna temperature
 $\mathrm{T_A}^*$ was converted to main beam temperature $\mathrm{T_{mb}}$ through 
 $\mathrm{T_{mb}}=\mathrm{T_A}^*/\eta_{\mathrm{mb}} $ where $\eta_{\mathrm{mb}}$ is the main beam efficiency, taken 
to be 0.38 based on November
 2004 observations of the Saturn and 3C279.  
The typical r.m.s noise ranged from 10 to 20 mK, after reduction.  The spectra were then binned into
 velocities which best showed
 the emission features (typically 10 km/s or 15 km/s).  The integrated intensity $\rm{I_{CO}}$ is calculated by
$\rm{I_{CO}} = \int \mathrm{T_{mb}} dv$, were the range of velocity for integration was chosen with reference to spectra from
the FCRAO calatog (\cite{young95}, or Kenney and Young (1988)). 
The errors for the line intensities were calculated as 
\begin{equation}
\delta \mathrm{I_{CO}}= \sigma \sqrt{\Delta V_{CO} \delta V} \quad [\mathrm{K\ km\ s^{-1}}]
\label{eqerror}
\end{equation}
where $\sigma$ is the r.m.s. noise in $\mathrm{T_{mb}}$, $\Delta \mathrm{V_{CO}}$ the full line width,
and  $\delta \mathrm{V}$ the velocity resolution (5, 10, 15, or 20 $\mathrm{km\ s^{-1}}$).

For spectra with signal to noise ratio of less than 3, the upper limits were calculated by 
\begin{equation}
\mathrm{I_{CO}} \le 3 \sigma \sqrt{\Delta V_{CO} \delta V} \quad [\mathrm{K\ km\ s^{-1}}]
\end{equation}
For those galaxies which were detected in the FCRAO survey but not ours, the same $\Delta V_{CO}$ was assumed.  For galaxies which were 
detected in neither sample, an arbitraray $\Delta V_{CO}$ of 100 km/s was assumed.

The observational errors presented in the following section are only for the r.m.s. errors of the baseline.  In general, other sources
of error (pointing error, baseline subtraction and calibration) which are difficult to quantify, are dominant sources of
 error for millimeter observations.
Three galaxies (NGC\ 6951, NGC\ 6946 and IC\ 342) were observed both in this study and the survey by \cite{NN01} at the same telescope.
The integrated intensities of NGC\ 6951 and IC\ 342 are consistent to within 30 \%.  For NGC\ 6946, \citet{NN01} give a value which is 
$\sim 50$ \% of this study, but may be attributed to the bad observing condition in our study (system temperature 1400 K).  For most of the
samples, therefore, readers should assume a typical error of $\sim 30$ \% for the obtained integrated intensities.

 \section{Results}
We show the spectra of the observed galaxies in figure 1 (detected galaxies followed by tentative detections and non-detections).  Of 
the 68 galaxies observed,
 54 were detected at a signal to noise ratio (S/N) of more than
5, and 60 galaxies with S/N more than 3.  All analysis in this paper
will be limited towards galaxies with
S/N over 5.  

\clearpage
\begin{figure}[h]
  \begin{center}
  \begin{tabular}{cc}
  \includegraphics[width=7cm,height=6cm]{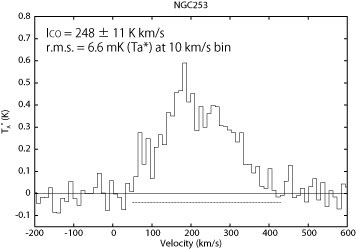} \label{N253}
  \includegraphics[width=7cm,height=6cm]{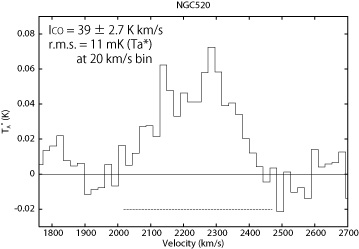} \label{N520} \\
  \includegraphics[width=7cm,height=6cm]{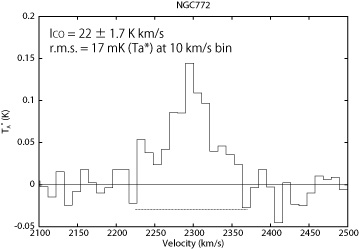} \label{N772}
  \includegraphics[width=7cm,height=6cm]{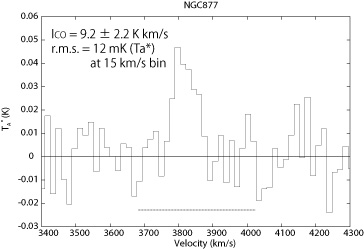} \label{N877} \\
  \includegraphics[width=7cm,height=6cm]{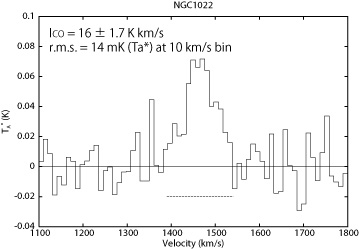} \label{N1022}
  \includegraphics[width=7cm,height=6cm]{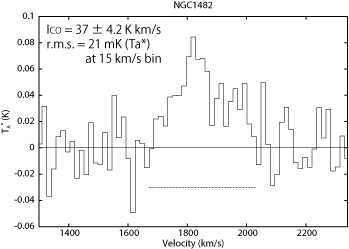} \label{N1482}
\end{tabular}
\label{spectra1}
\caption{Spectra of the galaxies observed at NRO 45m telescope, with S/N over 5.
   Absicca is the heliocentric velocity.  Emission is indicated by the
   dotted horizontal line, selected with reference to \cite{young95} or Kenney and Young (1988).}
     \end{center}
\end{figure}

\begin{figure}[h]
  \begin{center}
  \begin{tabular}{cc}
  \includegraphics[width=7cm,height=6cm]{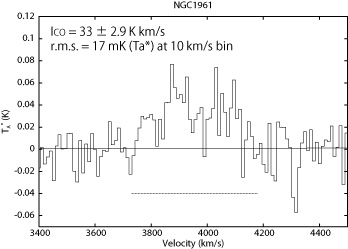} \label{N1961} 
  \includegraphics[width=7cm,height=6cm]{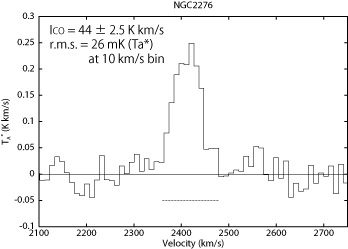} \label{N2276}\\
  \includegraphics[width=7cm,height=6cm]{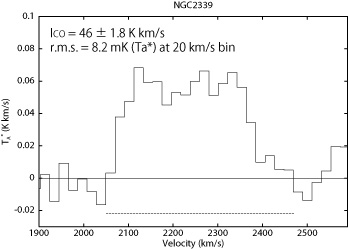} \label{N2339}
  \includegraphics[width=7cm,height=6cm]{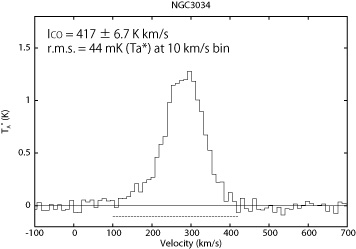} \label{N3034}\\
  \includegraphics[width=7cm,height=6cm]{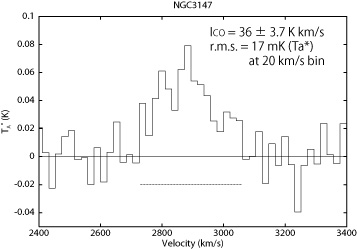} \label{N3147}
  \includegraphics[width=7cm,height=6cm]{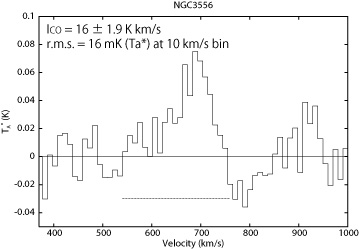} \label{N3556}
\end{tabular}
\label{spectra2}
     \end{center}
\end{figure}

\begin{figure}[h]
  \begin{center}
  \begin{tabular}{cc}
  \includegraphics[width=7cm,height=6cm]{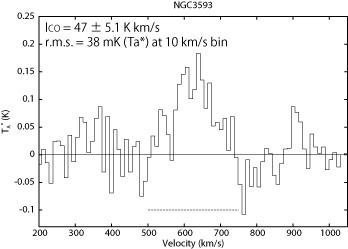} \label{N3593}
  \includegraphics[width=7cm,height=6cm]{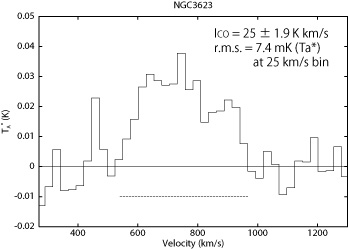} \label{N3623}\\
  \includegraphics[width=7cm,height=6cm]{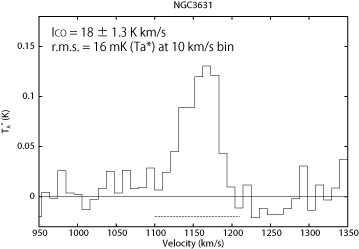} \label{N3631}
  \includegraphics[width=7cm,height=6cm]{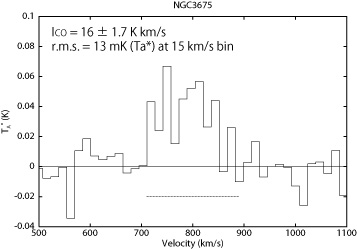} \label{N3675}\\
  \includegraphics[width=7cm,height=6cm]{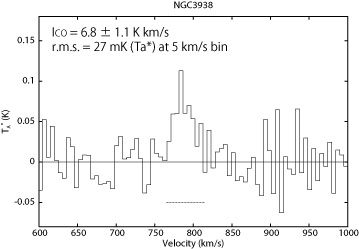} \label{N3938}
  \includegraphics[width=7cm,height=6cm]{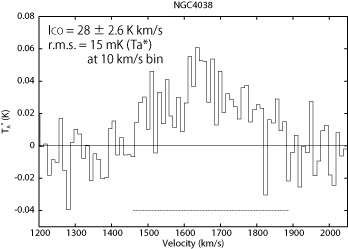} \label{N4038}
\end{tabular}
\label{spectra3}
     \end{center}
\end{figure}

\begin{figure}[h]
  \begin{center}
  \begin{tabular}{cc}
  \includegraphics[width=7cm,height=6cm]{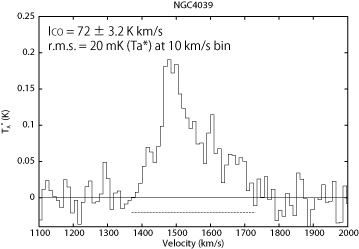} \label{N4039}
  \includegraphics[width=7cm,height=6cm]{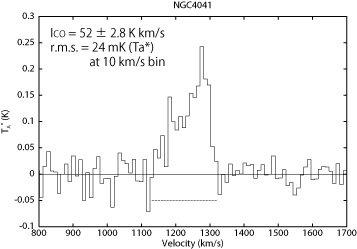} \label{N4041}\\
  \includegraphics[width=7cm,height=6cm]{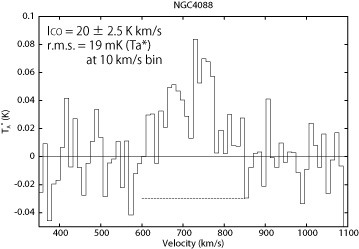} \label{N4088}
  \includegraphics[width=7cm,height=6cm]{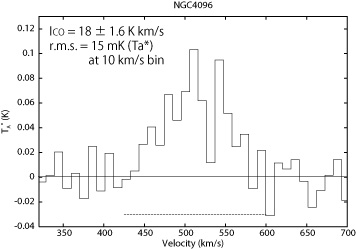} \label{N4096}\\
  \includegraphics[width=7cm,height=6cm]{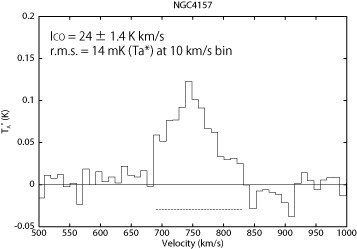} \label{N4157}
  \includegraphics[width=7cm,height=6cm]{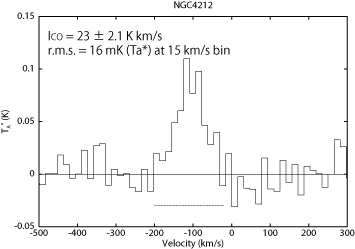} \label{N4212}
\end{tabular}
\label{spectra4}
     \end{center}
\end{figure}

\begin{figure}[h]
  \begin{center}
  \begin{tabular}{cc}
  \includegraphics[width=7cm,height=6cm]{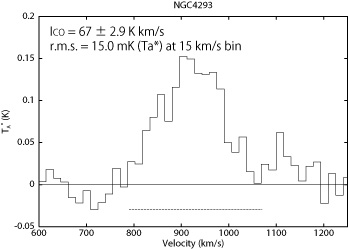} \label{N4293}
  \includegraphics[width=7cm,height=6cm]{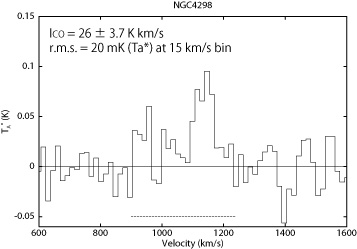} \label{N4298}\\
  \includegraphics[width=7cm,height=6cm]{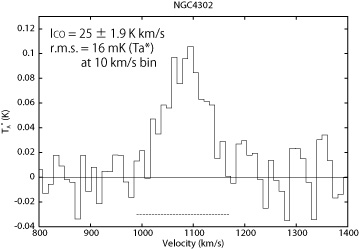} \label{N4302}
  \includegraphics[width=7cm,height=6cm]{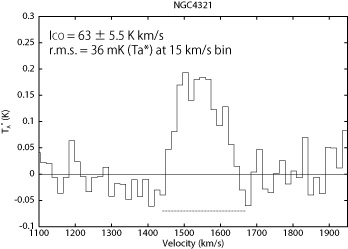} \label{N4321}\\
  \includegraphics[width=7cm,height=6cm]{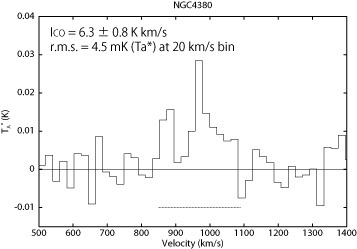} \label{N4380}
  \includegraphics[width=7cm,height=6cm]{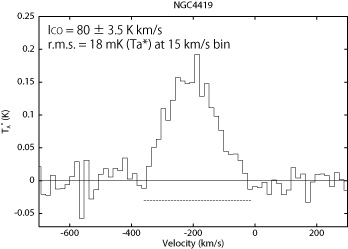} \label{N4419}
\end{tabular}
\label{spectra5}
     \end{center}
\end{figure}

\begin{figure}[h]
  \begin{center}
  \begin{tabular}{cc}
  \includegraphics[width=7cm,height=6cm]{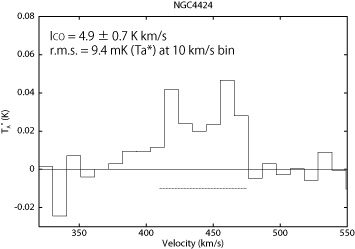} \label{N4424}
  \includegraphics[width=7cm,height=6cm]{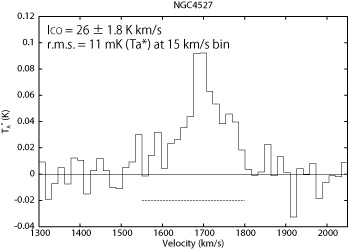} \label{N4527}\\
  \includegraphics[width=7cm,height=6cm]{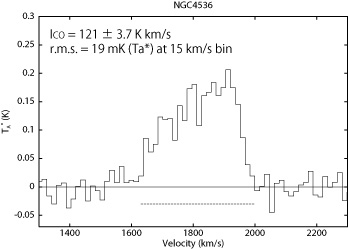} \label{N4536}
  \includegraphics[width=7cm,height=6cm]{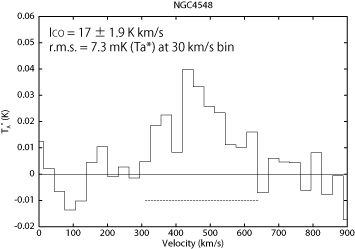} \label{N4548}\\
  \includegraphics[width=7cm,height=6cm]{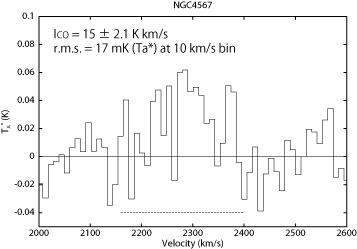} \label{N4567}
  \includegraphics[width=7cm,height=6cm]{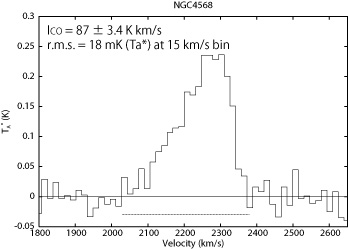} \label{N4568}
\end{tabular}
\label{spectra6}
     \end{center}
\end{figure}

 \begin{figure}[h]
  \begin{center}
  \begin{tabular}{cc}
  \includegraphics[width=7cm,height=6cm]{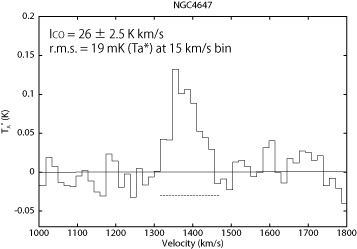} \label{N4647}
  \includegraphics[width=7cm,height=6cm]{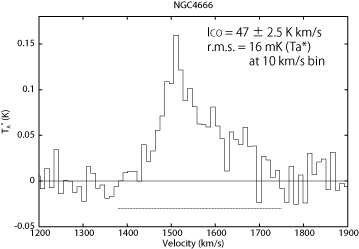} \label{N4666}\\
  \includegraphics[width=7cm,height=6cm]{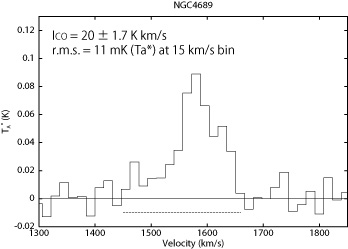} \label{N4689}
  \includegraphics[width=7cm,height=6cm]{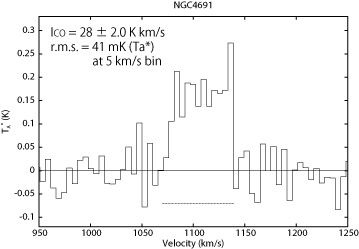} \label{N4691}\\
  \includegraphics[width=7cm,height=6cm]{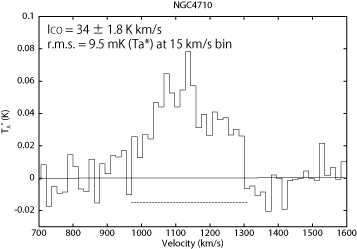} \label{N4710}
  \includegraphics[width=7cm,height=6cm]{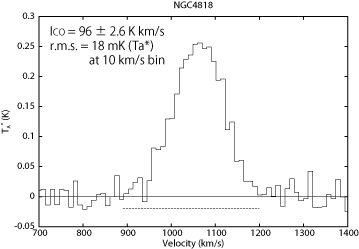} \label{N4818}

\end{tabular}
\label{spectra7}
     \end{center}
\end{figure}

\begin{figure}[h]
  \begin{center}
  \begin{tabular}{cc}
  \includegraphics[width=7cm,height=6cm]{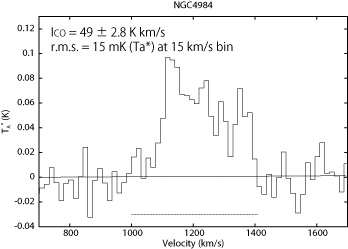} \label{N4984}
  \includegraphics[width=7cm,height=6cm]{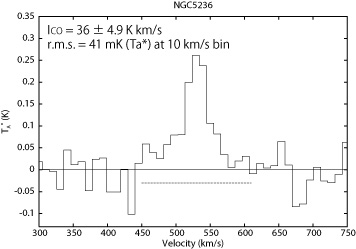} \label{N5236}\\
  \includegraphics[width=7cm,height=6cm]{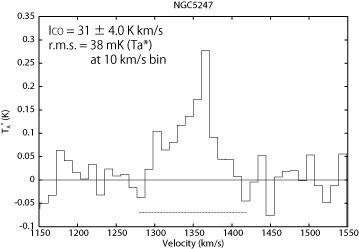} \label{N5247}
  \includegraphics[width=7cm,height=6cm]{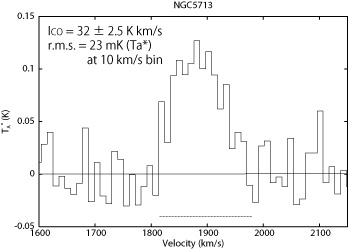} \label{N5713}\\
  \includegraphics[width=7cm,height=6cm]{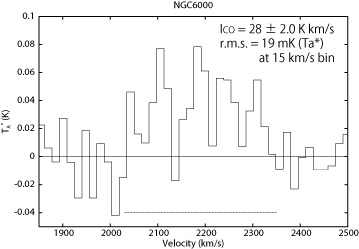} \label{N6000}
  \includegraphics[width=7cm,height=6cm]{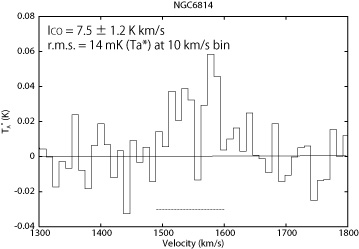} \label{N6814}
\end{tabular}
\label{spectra8}
     \end{center}
\end{figure}

\begin{figure}[h]
  \begin{center}
  \begin{tabular}{cc}
  \includegraphics[width=7cm,height=6cm]{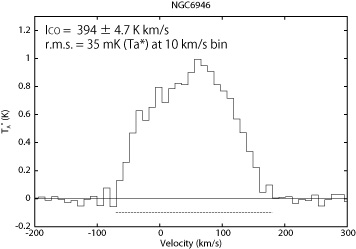} \label{N6946}
  \includegraphics[width=7cm,height=6cm]{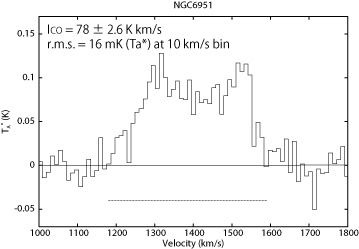} \label{N6951}\\
  \includegraphics[width=7cm,height=6cm]{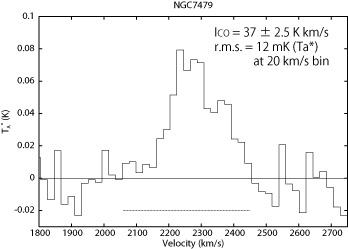} \label{N7479}
  \includegraphics[width=7cm,height=6cm]{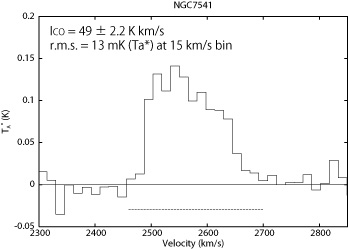} \label{N7541}\\
  \includegraphics[width=7cm,height=6cm]{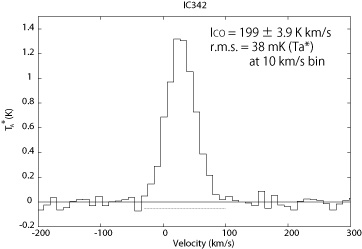} \label{IC342}
  \includegraphics[width=7cm,height=6cm]{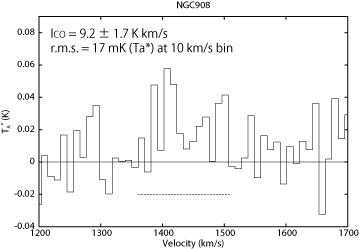} \label{N908}
\end{tabular}
\label{spectra9}
     \end{center}
\end{figure}

\begin{figure}[h]
  \begin{center}
  \begin{tabular}{cc}
  \includegraphics[width=7cm,height=6cm]{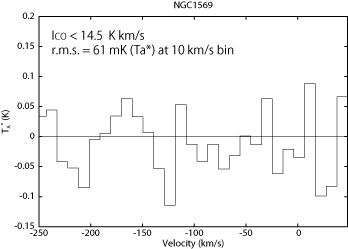} \label{N1569}
  \includegraphics[width=7cm,height=6cm]{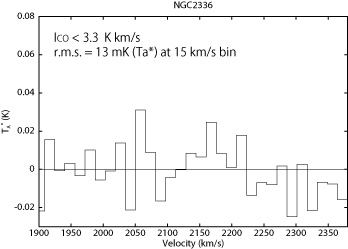} \label{N2336}\\
  \includegraphics[width=7cm,height=6cm]{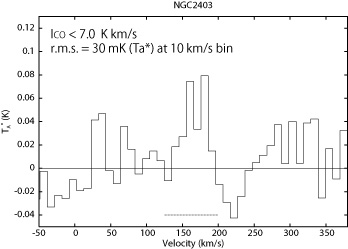} \label{N2403}
  \includegraphics[width=7cm,height=6cm]{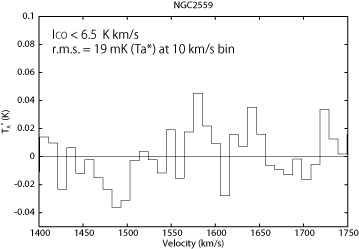} \label{N2559}\\
  \includegraphics[width=7cm,height=6cm]{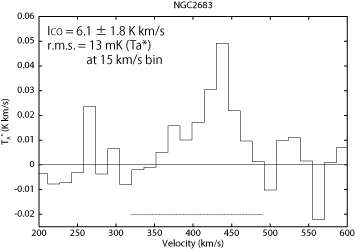} \label{N2683}
  \includegraphics[width=7cm,height=6cm]{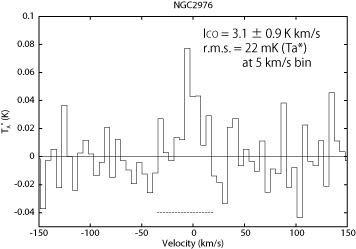} \label{N2976}
\end{tabular}
\label{spectra2_1}
\caption{Same as figure 1, for galaxies below S/N of 5.  Integrated intensity is shown only for those 
over S/N of 3, and upper limits for S/N less than 3.}
     \end{center}
\end{figure}

\begin{figure}[h]
  \begin{center}
  \begin{tabular}{cc}
  \includegraphics[width=7cm,height=6cm]{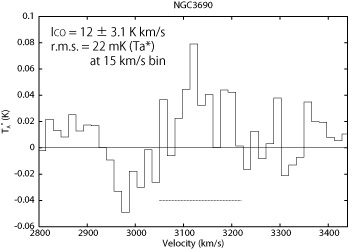} \label{N3690} 
  \includegraphics[width=7cm,height=6cm]{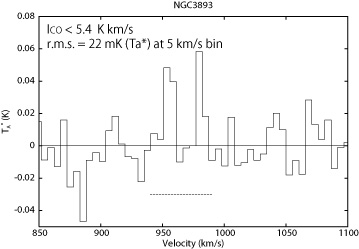} \label{N3893}\\
  \includegraphics[width=7cm,height=6cm]{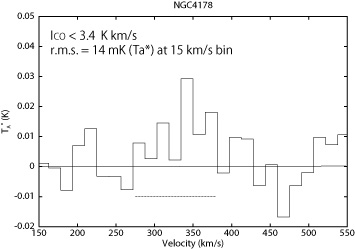} \label{N4178}
  \includegraphics[width=7cm,height=6cm]{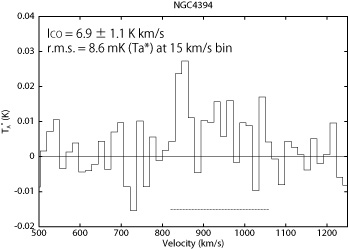} \label{N4394}\\
  \includegraphics[width=7cm,height=6cm]{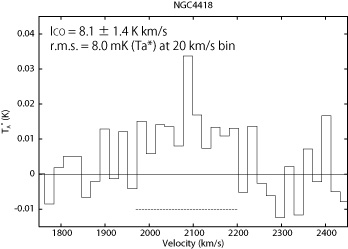} \label{N4418}
  \includegraphics[width=7cm,height=6cm]{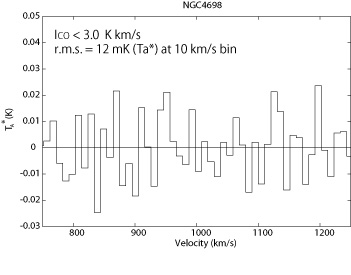} \label{N4698}
\end{tabular}
\label{spectra2_2}
\end{center}
\end{figure}

\begin{figure}[h]
  \begin{center}
  \begin{tabular}{cc}
  \includegraphics[width=7cm,height=6cm]{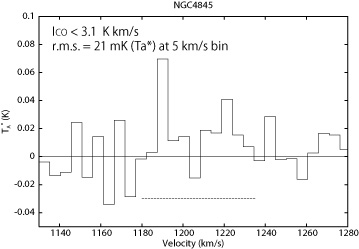} \label{N4845}
  \includegraphics[width=7cm,height=6cm]{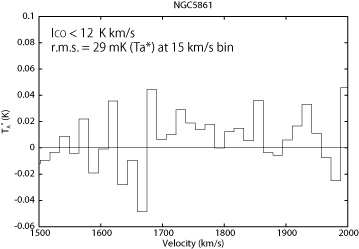} \label{N5861}
\end{tabular}
\label{spectra2_3}
\end{center}
\end{figure}
\clearpage

  \section{Other Data and Consistency}
In addition to the data gathered in this study at the NRO 45m telescope,
other galaxies which were observed in other previous studies either at
NRO 45m or the IRAM 30m telescope were compiled.  The IRAM 30m telescope
has a beamwidth of 22$^{\prime \prime}$, similar to the NRO45m
telescope.  After compilation, we have a total sample of 166 galaxies
observed in CO at the central region.
Furthermore, 117 of these were observed by Young et al. (1995) or Kenney \& Young (1988) with
45$^{\prime \prime}$ resolution.

Table \ref{tbl-1} lists the compiled data, along with the derived integrated intensities from this study.

\clearpage

\begin{deluxetable}{lllllll}
\footnotesize
\tablecolumns{7}
\tablecaption{Compiled Galaxy Samples \label{tbl-1}}
\tablewidth{0pt}
\tablehead{
\colhead{Galaxy} & \colhead{$\mathrm{I_{CO}}$}& \colhead{Ref.}& \colhead{Distance} & \colhead{$\mathrm{I_{CO}^{FC}}$}
& \colhead{Hubble type} & \colhead{Morphology} \\
\colhead{} & \colhead{K km/s} & & \colhead{Mpc} & \colhead{K km/s} &
 \colhead{HyperLEDA} & \colhead{NED} \\
\colhead{(1)} & \colhead{(2)} & \colhead{(3)} & \colhead{(4)} &
 \colhead{(5)} & \colhead{(6)} & \colhead{(7)} }
\startdata
 NGC\ 157  &  23$\pm$0.9   &2   &23.3    & 9.02   & 4.0 & SAB(rs)bc \nl
 NGC\ 253  & 248$\pm$11    &1   &2.2	& 467    & 5.1 & SAB(s)c   \nl
 NGC\ 278  &  18$\pm$0.8   &2,6 &11.8	& 14.2   & 2.9 & SAB(rs)b  \nl
 NGC\ 337a & \textless0.46 &5   &13.3*   & 35.5   & 7.9 & SAB(s)dm  \nl
 %NGC\ 428  & \textless0.74 &5   &13.3	& -      & 8.6 & SAB(s)m   \nl
 NGC\ 520  &  39$\pm$2.7   &1   &20.2    & 30.2   & 0.8 & Pec       \nl
 NGC\ 628  & 4.1$\pm$0.3   &2,6 &10.6 	& 4.15   & 5.2 & SA(s)c    \nl
 NGC\ 660  &  99$\pm$1.3   &2   &13.1	& 38.2   & 1.3 & SB(s)a pec\nl
 NGC\ 772  &  22$\pm$1.7   &1   &34.1    & 16.1   & 3.0 & SA(s)b    \nl
 NGC\ 864  &   6$\pm$0.9   &2   &21.8	& 1.31   & 5.1 & SAB(rs)c  \nl
 NGC\ 877  &  10$\pm$1.6   &1   &54.9    & 12.9   & 4.8 & SAB(rs)bc \nl
 NGC\ 891  &  96$\pm$5     &2   &9.4	& 26.2   & 3.0 & SA(s)b sp \nl
 NGC\ 908  & 9.2$\pm$1.7   &1   &19.6	& 18.9   & 5.1 & SA(s)c    \nl
 NGC\ 925  & 1.9$\pm$0.5   &2   &9.5	& 1.18   & 7.0 & SAB(s)d   \nl
 NGC\ 1022 &  16$\pm$1.7   &1   &20.1	& 9.67   & 1.1 & SB(s)a    \nl
 NGC\ 1042 & 2.9$\pm$0.3   &2   &19.1**  & -      & 6.0 & SAB(rs)cd \nl
 NGC\ 1055 &  46$\pm$1.6   &2   &14.3	& 41.1   & 3.2 & SBb sp    \nl
 NGC\ 1068 & 218$\pm$7     &3   &15.1	& 99.8   & 3.0 & SA(rs)b   \nl
 NGC\ 1084 &  31$\pm$2.2   &2   &18.7	& 11.5   & 4.9 & SA(s)c    \nl
 NGC\ 1087 &  15$\pm$1     &2   &24.6	& 6.86   & 5.2 & SAB(rs)c  \nl
 NGC\ 1482 &  37$\pm$4.2   &1   &20.5	& 20.0   & -0.9& SA0+ pec sp \nl
 NGC\ 1569 & \textless14.5  &1   &3.1    & 1.82   & 9.6 & IBm       \nl
 NGC\ 1637 &  13$\pm$0.7   &2   &9.5**	& -      & 5.0 & SAB(rs)c  \nl
 NGC\ 1961 &  33$\pm$2.9   &1   &54.1    & 29.5   & 4.2 & SAB(rs)c  \nl
 NGC\ 2146 & 118$\pm$1.9   &2   &13.7	& 68.4   & 2.3 & SB(s)ab pec \nl 
 NGC\ 2276 &  44$\pm$2.5   &1   &34.4	& 11.4   & 4.0 & SAB(rs)c  \nl
 NGC\ 2336 & \textless3.3 &1   &31.9	& \textless1.01 & 5.4& SAB(r)bc \nl
 NGC\ 2339 &  46$\pm$1.8   &1   &31.1	& 16.8  & 4.0  & SAB(rs)bc \nl
 NGC\ 2403 & \textless7.0  &1   &2.2	& 4.91  & 6.0  & SAB(s)cd  \nl
 %%NGC\ 2552 & \textless0.79 &5   &9.2*	&       & 4.5  & SA(s)m    \nl
 NGC\ 2559 & \textless6.5  &1   &17.3	& 32.1  & 9.0  & SB(s)bc pec \nl
 NGC\ 2681 &  30$\pm$0.8   &2   &10.1	& 7.6   & 3.1  & SAB(rs)0/a \nl
 NGC\ 2683 & 6.1$\pm$1.8   &1   &3.2	& 5.07  & 0.4  & SA(rs)b   \nl
 NGC\ 2715 & 7.6$\pm$1.5   &2   &20.3**  & -     & 5.2  & SAB(rs)c  \nl
 NGC\ 2805 & 2.3$\pm$0.1   &5   &26.2*	& -     & 6.9  & SAB(rs)d  \nl
 NGC\ 2820 & 9.8$\pm$1.1   &2   &22.9**  & -     & 5.4  & SB(s)c pec sp\nl
 NGC\ 2841 &   6$\pm$0.6   &2   &9.3	& 1.46  & 3.0  & SA(r)b    \nl
 NGC\ 2903 & 119$\pm$2.3   &6   &6.2 	& 24    & 4.0  & SB(s)d    \nl
 NGC\ 2964 &  25$\pm$0.9   &2   &16.8	& 11.4  & 4.1  & SAB(r)bc  \nl
 NGC\ 2976 & 3.1$\pm$0.9*** &1   &2.3	& 2.07  & 5.3  & SAc pec   \nl
 NGC\ 2985 &  12$\pm$2.9   &2   &19.1	& 9.55  & 2.3  & SA(rs)ab  \nl
 NGC\ 3034 & 417$\pm$6.7   &1   &2.2	& 289   & 8.0  & I0        \nl
 NGC\ 3079 & 212$\pm$2.7   &2   &16.1	& 69.3  & 6.6  & SB(s)c    \nl
 NGC\ 3147 &  36$\pm$3.7   &1   &38.4	& 10.5  & 3.9  & SA(rs)bc  \nl
 NGC\ 3184 & 7.8$\pm$1     &6   &7.9	& 5.8   & 5.0  & SAB(rs)cd \nl
 NGC\ 3187 & 3.5$\pm$0.6   &2   &20.0**  & -     & 5.9  & SB(s)c pec \nl
 NGC\ 3198 &  11$\pm$0.5   &2   &9.3**	& -     & 5.2  & SB(rs)c   \nl
 %%NGC\ 3206 & \textless0.52 &5   &25.8*	& -     & 6.0  & SB(s)cd   \nl
 NGC\ 3227 &  54$\pm$3.7   &2   &14.3**  & -     & 1.4  & SAB(s) pec \nl
 NGC\ 3310 & 3.6$\pm$0.9   &2   &14.2	& 2.15  & 4.0  & SAB(r)bc pec \nl
 NGC\ 3344 & 4.3$\pm$0.4   &2   &6.9	& 4.69  & 4.0  & SAB(r)bc  \nl
 NGC\ 3346 & 3.6$\pm$0.23  &5   &17.5*	& -     & 6.0  & SB(rs)cd  \nl
 NGC\ 3351 &  17$\pm$1.9   &2   &9.0	& 14.6  & 3.0  & SB(r)b    \nl
 NGC\ 3359 &   3$\pm$1     &2   &15.0	& 4.95  & 5.2  & SB(rs)c   \nl
 NGC\ 3368 &  35$\pm$1     &2   &10.3	& 24.4  & 1.8  & SAB(rs)ab \nl
 NGC\ 3423 & 2.7$\pm$0.24  &5   &13.6*	& -     & 6.0  & SA(s)cd   \nl
 NGC\ 3445 &0.89$\pm$0.1   &5   &30.0*	& -     & 8.9  & SAB(s)m   \nl
 NGC\ 3486 &   7$\pm$1     &2   &9.0	& 0.76  & 5.2  & SAB(r)c   \nl
 NGC\ 3504 & 103$\pm$3.1   &6   &19.7	& 13.7  & 2.1  & SAB(s)ab  \nl
 NGC\ 3521 &  17$\pm$1.1   &6   &8.5	& 22.7  & 4.0  & SAB(rs)bc \nl
 NGC\ 3556 &  16$\pm$1.9   &1   &10.3	& 14.4  & 6.0  & SB(s)cd   \nl
 NGC\ 3593 &  47$\pm$5.1   &1   &7.3	& 18.6  & -0.4 & SA(s)0/a  \nl
 NGC\ 3623 &  25$\pm$1.9   &1   &4.5     & \textless1.90  &1.0 & SAB(rs)a \nl
 NGC\ 3627 &  90$\pm$4     &2   &4.5	& 34    &3.0   & SAB(rs)a  \nl
 NGC\ 3628 & 130$\pm$5.7   &2   &4.5	& 69.5  &3.1   & SAb pec sp\nl
 NGC\ 3631 &  18$\pm$1.3   &1   &16.6	& 7.66  &5.2   & SA(s)c    \nl
 NGC\ 3675 &  16$\pm$1.7   &1   &9.8	& 12.2  &3.0   & SA(s)b    \nl
 NGC\ 3690 &  12$\pm$3.1   &1   &41.4    & 19.5  &8.7   & IBm pec\nl
 NGC\ 3782 &0.68$\pm$0.11  &5   &12.6*	& -     &6.6   & SAB(s)cd  \nl
 NGC\ 3810 &  24$\pm$2     &1,6 &11.5	& 14.3  &5.2   & SA(rs)c   \nl
 NGC\ 3893 & \textless5.4 &1   &13.8	& 9.72  &5.1   & SAB(rs)c  \nl
 NGC\ 3906 &0.37$\pm$0.12  &5   &15.6*	& -     & 6.8  & SB(s)d    \nl
 NGC\ 3913 & 1.9$\pm$0.22  &5   &15.9*	& -     & 6.6  & SA(rs)d   \nl
 NGC\ 3938 & 6.8$\pm$1.1   &1   &11.2	& 8.86  & 5.1  & SA(s)c    \nl
 NGC\ 4038 &  28$\pm$2.6   &1   &19.3	& 29.3  & 8.9  & SB(s)m pec\nl
 NGC\ 4039 &  72$\pm$3.2   &1   &19.3	& 29.6  & 8.9  & SA(s)m pec\nl
 NGC\ 4041 &  52$\pm$2.8   &1   &17.6    & 20.4  & 4.0  & SA(rs)bc  \nl
 NGC\ 4088 &  20$\pm$2.5   &1   &10.9	& 15.6  & 4.8  & SAB(rs)bc \nl
 NGC\ 4096 &  18$\pm$1.6   &1   &7.5	& 5.07  & 5.3  & SAB(rs)c  \nl
 NGC\ 4102 & 115$\pm$9     &6   &13.1	& 17.5  & 3.0  & SAB(s)b   \nl
 NGC\ 4157 &  24$\pm$1.4    &1   &12.1    & 21.9  & 3.3  & SAB(s)b sp\nl
 NGC\ 4178 & \textless3.4   &1   &16.1	& 1.64  & 7.1  & SB(rs)dm  \nl
 NGC\ 4204 &0.50$\pm$0.14  &5   &12.9*	& -     & 7.9  & SB(s)dm   \nl
 NGC\ 4212 &  23$\pm$2.1   &1   &16.1	& 7.46  & 4.9  & SAc       \nl
 %%NGC\ 4242 & \textless0.90 &5   &9.8*	& -     & 8.0  & SAB(s)dm  \nl
 NGC\ 4254 &  47$\pm$7.4   &6   &16.1	& 19.3  & 5.2  & SA(s)c    \nl
 NGC\ 4274 &  10$\pm$0.4   &2   &11.9**  & -     & 1.7  & SB(r)ab   \nl
 NGC\ 4293 &  67$\pm$1.1   &1   &16.1	& 11.5  & 0.3  & SB(s)0/a  \nl
 NGC\ 4298 &  26$\pm$3.7   &1   &16.1	& 10.9  & 5.2  & SA(rs)c   \nl
 NGC\ 4299 &0.48$\pm$0.14  &5   &16.1	& 1.11  & 8.3  & SAB(s)dm  \nl
 NGC\ 4302 &  25$\pm$1.9   &1   &16.1	& 8.91  & 5.4  & Sc sp     \nl
 NGC\ 4303 &  69$\pm$2.1   &6   &16.1	& 21.6  & 4.0  & SAB(rs)bc \nl
 NGC\ 4314 &  24$\pm$0.3   &2   &10.9**  & -     & 1.0  & SB(rs)a   \nl
 NGC\ 4321 &  63$\pm$5.5   &1,2 &16.1	& 27.3  & 4.0  & SAB(s)bc  \nl
 NGC\ 4380 & 6.3$\pm$0.8  &1   &16.1    & \textless0.1  &2.3 & SA(rs)b \nl
 NGC\ 4394 & 6.9$\pm$1.1   &1   &16.1	& 3.27  & 8.8  & SB(r)b    \nl
 %%NGC\ 4395 & \textless0.58 &5   &6.6*	& -     & 2.9  & SA(s)m    \nl
 NGC\ 4411 &0.88$\pm$0.10  &5   &17.8*	& -     & 5.4  & SB(rs)c   \nl
 NGC\ 4414 &  79$\pm$5.5   &2,6 &9.6	& 31.6  & 5.1  & SA(rs)c   \nl
 NGC\ 4416 &5.08$\pm$0.19  &5   &19.3*	& -     & 5.9  & SB(rs)cd  \nl
 NGC\ 4418 & 8.1$\pm$1.4   &1   &25.5	& 4.91  & 1.0  & SAB(s)a   \nl
 NGC\ 4419 &  80$\pm$3.5   &1   &16.1	& 18.6  & 1.1  & SB(s)a    \nl
 NGC\ 4424 & 4.9$\pm$0.7   &1   &16.1    & 1.45  & 3.0  & SB(s)a    \nl
 NGC\ 4438 &  70$\pm$7     &4   &16.1    & 5.27  & 0.7  & SA(s)0/a pec \nl
 NGC\ 4487 &3.23$\pm$0.15  &5   &13.6*	& -     & 5.9  & SAB(rs)cd  \nl
 NGC\ 4496A&2.17$\pm$0.18  &5   &23.6*	& -     & 7.5  & SB(rs)m   \nl
 NGC\ 4501 &  51$\pm$2.5   &6   &16.1	& 21.8  & 3.4  & SA(rs)b   \nl
 %%NGC\ 4517A &\textless0.65 &5   &20.7*	& -     & 7.8  & SB(rs)dm  \nl
 NGC\ 4527 &  26$\pm$1.8   &1   &16.1	& 28.7  & 4.0  & SAB(s)bc  \nl
 %%NGC\ 4534 & \textless0.42 &5   &13.3*	& -     & 7.7  & SA(s)dm   \nl
 NGC\ 4535 &  42$\pm$1.2   &6   &16.1	& 8.55  & 5.0  & SAB(s)c   \nl
 NGC\ 4536 & 121$\pm$3.7   &1   &16.1	& 18.4  & 4.2  & SAB(rs)bc \nl
 NGC\ 4540 &5.57$\pm$0.19  &5   &16.1	& 1.50  & 6.1  & SAB(rs)cd \nl
 NGC\ 4548 &  17$\pm$1.9   &1   &16.1	& 6.73  & 3.1  & SBb(rs)   \nl
 NGC\ 4565 &  12$\pm$1.2   &2   &15.7**  & -     & 3.2  & SA(s)b sp \nl
 NGC\ 4567 &  15$\pm$2.1   &1   &16.1    & 8.55  & 4.0  & SA(rs)bc  \nl
 NGC\ 4568 &  87$\pm$3.4   &1   &16.1	& 19.1  & 4.1  & SA(rs)bc  \nl
 NGC\ 4569 & 134$\pm$3.2   &6   &16.1	& 26.6  & 2.4  & SAB(rs)ab \nl
 NGC\ 4618 &0.72$\pm$0.17  &5   &10.0*	& -     & 8.6  & SB(rs)m   \nl
 NGC\ 4625 &3.78$\pm$0.11  &5   &10.9*	& -     & 8.8  & SAB(rm)m pec\nl
 NGC\ 4647 &  26$\pm$2.5   &1   &16.1	& 9.09  & 5.2  & SAB(rs)c  \nl
 NGC\ 4651 &   6$\pm$1     &2   &16.1	& 7.64  & 5.2  & SA(rs)c   \nl
 NGC\ 4654 &  24$\pm$0.9   &2   &16.1	& 6.0   & 5.9  & SAB(rs)cd \nl
 NGC\ 4666 &  47$\pm$2.5   &1   &18.6	& 30.2  & 5.0  & SABc      \nl
 NGC\ 4688 &0.44$\pm$0.08  &5   &13.9*	& -     & 6.0  & SB(s)cd   \nl
 NGC\ 4689 &  20$\pm$1.7   &1   &16.1	& 6.55  & 4.7  & SA(rs)bc  \nl
 NGC\ 4691 &  28$\pm$2.0   &1   &13.1	& 4.8   & 0.4  & SB(s)0/a pec\nl
 NGC\ 4698 & \textless3.0 &1   &16.1	& \textless0.73& 1.4 & SA(s)ab \nl
 NGC\ 4701 & 2.9$\pm$0.22  &5   &10.3*	& -     & 5.8  & SA(s)cd   \nl
 NGC\ 4710 &  34$\pm$1.8   &1   &16.1	& 7.82  & -0.8 & SA(r)0 sp \nl
 NGC\ 4736 &  61$\pm$3.9   &6   &4.0	& 17.5  & 2.4  & SA(r)ab   \nl
 NGC\ 4775 &1.63$\pm$0.22  &5   &20.9*	& -     & 6.9  & SA(s)d    \nl
 NGC\ 4818 &  96$\pm$2.6   &1   &13.5	& 16.8  & 2.0  & SAB(rs)ab pec \nl
 NGC\ 4826 & 140$\pm$1.9   &6   &5.0	& 44.7  & 2.4  & SA(rs)ab  \nl
 NGC\ 4845 & \textless3.1  &1   &16.1	& 6.07  & 2.3  & SA(s)ab sp \nl
 NGC\ 4984 &  49$\pm$2.8   &1   &14.7	& 12.3  & -0.8 & SAB(rs)0  \nl
 NGC\ 5005 &  76$\pm$5.4   &2   &14.3	& 30.4  & 4.0  & SAB(rs)bc \nl
 NGC\ 5033 &  59$\pm$4     &2,6 &12.8	& 13.1  & 5.2  & SA(s)c    \nl
 NGC\ 5112 & 1.7$\pm$0.4   &2   &13.4**  & -     & 5.8  & SB(rs)cd  \nl
 NGC\ 5236 &  36$\pm$4.9   &1   &5.9	& 88.7  & 5.0  & SAB(s)c   \nl
 NGC\ 5247 &  31$\pm$4     &1   &20.1	& 7.67  & 4.0  & SA(s)bc   \nl
 NGC\ 5248 &  67$\pm$2.3   &6   &14.7	& 19.5  & 4.1  & SB(rs)bc  \nl
 NGC\ 5364 &   5$\pm$1     &2   &18.0	& 1.00  & 4.0  & SA(rs)bc pec\nl
 %%NGC\ 5477 & \textless0.43 &5   &7.6*	& -     & 8.7  & SA(s)m    \nl
 %%NGC\ 5584 & \textless0.50 &5   &22.6*	& -     & 5.9  & SAB(rs)cd \nl
 NGC\ 5668 &1.78$\pm$0.26  &5   &22.2*	& -     & 6.0  & SA(s)d    \nl
 NGC\ 5669 &2.28$\pm$0.27  &5   &19.8*   & -     & 6.9  & SAB(rs)cd \nl
 NGC\ 5713 &  32$\pm$2.5   &1   &24.6	& 11.0  & 4.0  & SAB(rs)bc pec\nl
 NGC\ 5725 &1.12$\pm$0.24  &5   &22.8*	& -     & 7.0  & SB(s)d    \nl
 %%NGC\ 5789 & \textless0.74 &5   &26.7*	& -     & 7.8  & Sdm       \nl
 NGC\ 5861 & \textless12  &1   &24.1	& 10.2  & 5.0  & SAB(rs)c  \nl
 NGC\ 5907 &  32$\pm$1.6   &2   &10.4	& 7.49  & 5.4  & SA(s)c sp \nl
 NGC\ 5921 &   6$\pm$1.8   &2   &19.0**  & -     & 4.0  & SB(r)bc   \nl
 NGC\ 5964 &0.89$\pm$0.17  &5   &20.7*	& -     & 6.9  & SB(rs)d   \nl
 NGC\ 6000 &  28$\pm$2.0   &1   &27.1	& 39.1  & 3.9  & SB(s)bc   \nl
 NGC\ 6015 & 5.1$\pm$0.8   &2   &13.6**  & -     & 6.0  & SA(s)cd   \nl
 NGC\ 6217 &  19$\pm$1.6   &2   &21.1	& 10.2  & 4.0  & SB(rs)bc  \nl
 NGC\ 6384 & 7.6$\pm$0.7   &2   &24	& 3.27  & 3.6  & SAB(r)bc  \nl
 NGC\ 6503 &  11$\pm$1.1   &6   &4.2	& 3.78  & 5.9  & SA(s)cd   \nl
 NGC\ 6509 &5.97$\pm$0.23  &5   &25.7*	& -     & 6.6  & SBcd      \nl
 NGC\ 6643 &  14$\pm$1.6   &6   &23.1	& 13.8  & 5.2  & SA(rs)c   \nl
 NGC\ 6814 & 7.5$\pm$1.2   &1   &21.1	& 4.71  & 4.0  & SAB(rs)bc \nl
 NGC\ 6946 & 394$\pm$4.7   &1,6 &6.7	& 79.8  & 5.9  & SAB(rs)cd \nl
 NGC\ 6951 &  78$\pm$2.6   &1,6 &21.7	& 15.2  & 3.9  & SAB(rs)bc \nl
 NGC\ 7217 &   7$\pm$1.4   &2,6 &16.3	& 4.13  & 2.5  & SA(r)ab   \nl
 NGC\ 7331 &  19$\pm$1.3   &2,6 &14.7	& 6.98  & 3.9  & SA(s)b    \nl
 NGC\ 7479 &  37$\pm$2.5   &1   &34.7    & 12.9  & 4.4  & SB(s)c    \nl
 NGC\ 7541 &  49$\pm$2.2   &1   &38.1    & 14.8  & 4.7  & SB(rs)bc pec\nl
 NGC\ 7640 & 2.3$\pm$0.3   &2   &8.5	& 1.33  & 5.3  & SB(s)c    \nl
 NGC\ 7741 & 1.1$\pm$0.3   &2,5 &11.7*	& -     & 6.0  & SB(s)cd   \nl
 IC\  342  & 199$\pm$3.9   &1,6 &3.0	& 65.5  & 5.9  & SAB(rs)cd \nl
 %%IC\  694  &  11$\pm$5     &1   &41.4	& 10.4  & 8.8  & SBm pec\nl
 %%IC\  750  &  58$\pm$3.6   &2   &9.6**	& -     & 2.2  & Sab sp    \nl
 UGC\ 3574 &1.16$\pm$0.18  &5   &21.9*	& -     & 5.9  & SA(s)cd   \nl
 %%UGC\ 3826 & \textless0.26 &5   &25.9*	& -     & 6.5  & SAB(s)d   \nl
 %%UGC\ 4499 & \textless0.44 &5   &11.7*	& -     & 7.9  & SABdm     \nl
 %%UGC\ 5015 & \textless0.71 &5   &24.0*	& -     & 7.7  & SABdm     \nl
 %%UGC\ 5288 & \textless0.53 &5   &7.5*	& -     & 8.0  & Sdm       \nl
 %%UGC\ 6931 & \textless0.50 &5   &19.3*	& -     & 9.0  & SBm       \nl
 UGC\ 8516 &1.20$\pm$0.19  &5   &15.4*	& -     & 5.9  & Scd       \nl
 %%UGC\ 12082& \textless0.46 &5   &13.0*	& -     & 8.7  & Sm        \nl
 %%UGC\ 12732& \textless 0.40&5   &11.6*	& -     & 8.7  & Sm        \nl
 MCG1-3-85\ &1.39$\pm$0.18 &5   &13.6*   & -     & 7.0  & SAB(rs)d  \nl
\enddata
\tablenotetext{(1)}{Galaxy name.}
\tablenotetext{(2)}{Integrated intensity of \co\Ja \, in
$\mathrm{T_{mb}}$.  *** for galaxies with bad pointing accuracy.}
\tablenotetext{(3)}{Reference of the CO intensity in column
 2. 1-This study, at NRO 45m.  2-\cite{braine} at IRAM 30m.
  3-\cite{planesas} at IRAM 30m. 4-\cite{combes} at IRAM 30m. 5-\cite{boeker}, at IRAM 30m.  6-\cite{NN01}, at NRO
 45m.  For galaxies outside of those observed in this study, only those which were detected are listed.}
\tablenotetext{(4)}{Distance to the galaxy in Mpc, from \cite{young95}
  unless otherwise noted;
 *:From \citet{boeker}. **:From \cite{braine}. ***:\cite{liu}. Values have been converted
 using 
$H_0=75\ \mathrm{km\ s^{-1}\ Mpc^{-1}}$.
  For members of the Virgo cluster, 16.1 Mpc is assumed from Cepheid
 calibrations (\cite{ferrarese}).}
\tablenotetext{(5)}{Integrated intensity $\int \mathrm{T_{mb}}dv$ of the center from
the FCRAO sample by \cite{young95} or \cite{KY88} with a beamsize of 45$^{\prime \prime}$, converted from 
antenna temperature $\mathrm{T_{a}^*}$ to main beam temperature scale
using a main beam efficiency of $\eta_\mathrm{mb} = 0.55$ and $\mathrm{T_{a}^*} = \eta_\mathrm{mb} \mathrm{T_{mb}}$.}
\tablenotetext{(6)}{Hubble type, from the HyperLEDA database (Paturel et al. 2003).}
\tablenotetext{(7)}{Morphology, taken from NED except for NGC\ 3690, which is from RC2.}
\end{deluxetable}

\clearpage

\begin{figure}
  \begin{center}
  \includegraphics[width=15cm,height=15cm,keepaspectratio]{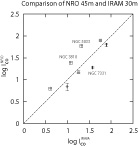}
  \end{center}
  \caption{Comparison of integrated CO intensities in $\mathrm{K\ km/s}$ for galaxies which
 were observed with both the IRAM 30m and NRO 45m telescope.  The galaxies
 are NGC\ 278, NGC\ 628, NGC\ 3810, NGC\ 4321, NGC\ 4414, NGC\ 5033,
 NGC\ 7217, NGC\ 7331.  The line expresses $\mathrm{I_{CO}}^\mathrm{NRO}=\mathrm{I_{CO}}^\mathrm{IRAM}$.} \label{comparison}
\end{figure}

Although the beamwidth of the IRAM 30m telescope at 22$^{\prime \prime}$ is similar to that of
 the NRO 45m telescope at 16$^{\prime \prime}$, it is important that 
CO detected at both telescopes can be compared on a same scale.  In order to confirm this point, figure
 \ref{comparison} plots the 
derived integrated intensity $\mathrm{I_{CO}}$ for galaxies which overlapped between observations at NRO 45m and IRAM 30m 
telescope.  A clear proportionality with a slope of order unity can be seen, despite
 the small number of samples.  The zero point of this relation is
 effectively 0, despite the difference in beam size, enabling a direct comparison
 of observed intensity.  

Three galaxies deviate from the relation in \ref{comparison} by more than a factor of two (NGC\ 3810, NGC\ 5033, and NGC\ 7331).
All of these galaxies are from \cite{NN01} and \cite{braine} for the NRO 45m and IRAM 30m measurements, respectively.
NGC\ 3810 has a starbursting nuclei, and its $\mathrm{I_{CO}}$ at a single beam away from its nuclei drops to nearly half
\cite{NN01}.  The observing coordinate in \cite{NN01} and \cite{braine} are offset by 4$^{\prime \prime}$, possibly explaining the 
excess of $\mathrm{I_{CO}}$ in the NRO 45m measurement.  The $\mathrm{I_{CO}}$ for NGC\ 5033 also drops by more than 50\% 
$11^{\prime \prime}$ from the center in \cite{NN01}, making the CO intensity consistent with that of \cite{braine}.  The offset in 
observing coordinate ($\sim 2^{\prime \prime}$) coupled with pointing errors may account for the discrepancy.  The observing coordinate for
NGC\ 7331 differ by $14^{\prime \prime}$.  Since the coordinate in \cite{braine} is offset from the true center, the difference in
$\mathrm{I_{CO}}$ may reflect the central CO hole which explains the smaller CO intensity in \cite{NN01}.
Nevertheless, it is important to bear in mind that when comparing CO intensities from the different telescopes, an error
of typically a factor 2 may infact be involved.

\begin{figure}
  \begin{center}
 \includegraphics[width=15cm,height=15cm,keepaspectratio]{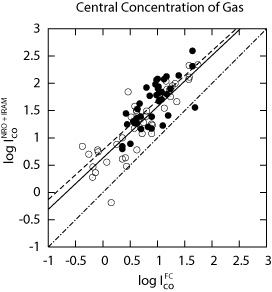}
  \end{center}
  \caption{The central concentration of gas.  Absissa is integrated CO
 intensity from the FCRAO survey with 45$^{\prime \prime}$ resolution, and the
 ordinate is same for telescopes IRAM 30m (open circles) and NRO 45m (filled
 circles) with 22$^{\prime \prime}$ and 16$^{\prime \prime}$ resolution, respectively.  The dashed line represents
a least squares fit to the NRO 45m sample, and the solid line IRAM 30m.  The dash-dotted line
 represents where the ratio of absissa and ordinate becomes unity.} \label{condensation}
\end{figure}

 A secondary check can be implemented by using the 45$^{\prime \prime}$ resolution data from the FCRAO data from Young et al. (1995).  The
 ratio of CO intensity at NRO 45m or IRAM 30m telescope to the FCRAO intensity expresses a measure of the central condensation of gas; if 
the angular resolution difference of the NRO 45m and IRAM 30m telescope has a significant effect on the derived intensity, this ratio 
should display a systematic difference between the two telescopes.  Figure \ref{condensation} plots the CO intensity observed at the 
FCRAO $\mathrm{I_{CO}^{FC}}$ versus the intensity observed either at NRO 45m or IRAM 30m telescope, including those only with more 
than a 3 $\sigma$ detection.  A clear correlation can be seen, and a least squares fit to the CO intensity from NRO45m
 telescope $\mathrm{I_{CO}^{NRO}}$ gives
\begin{equation}
\log \mathrm{I_{CO}}^\mathrm{NRO}=(0.91\pm0.14)\log \mathrm{I_{CO}}^\mathrm{FC}+(0.78\pm0.14),
\end{equation}

where as the same fit to IRAM30m telescope will give
\begin{equation}
\log \mathrm{I_{CO}}^\mathrm{IRAM}=(0.94\pm0.09)\log \mathrm{I_{CO}}^\mathrm{FC}+(0.63\pm0.08). 
\end{equation}
 Both fits are consistent within the range of error, and we conclude that the data from these two telescopes with different
 beamsizes can be compared on a same scale without correction.

\clearpage

\clearpage

  \section{Sample Properties}

\begin{figure}
  \begin{center}
  \includegraphics[width=10cm,height=8cm]{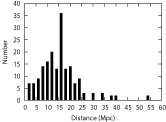}
  \end{center}
  \caption{Number distribution of the distance of the sample galaxies.
 The strong peak at 16Mpc are for members of the Virgo cluster.} \label{distancedistribution}
\end{figure}

Figure \ref{distancedistribution} shows the distance distribution of our compiled samples from
the NRO 45m and IRAM 30m dishes.  The samples are strongly peaked at
around 16 Mpc, where the Virgo cluster members lie.  The distances here
are taken from Young et al. (1995) scaled to $H_0 = 75\ \mathrm{km\
s^{-1}\ Mpc^{-1}}$ except for Virgo cluster members, where a more
recent measurement (\cite{ferrarese}) gives
16.1 Mpc using Cepheid variables.  The observing beam of 16$^{\prime \prime}$ and
22$^{\prime \prime}$ for 16 Mpc correspond to a linear scale of 1.2 kpc and 1.8 kpc,
respectively.  These both correspond to radii within 1 kpc of the center,
and considering that most of our samples lie around this distance, we will
simply refer hereafter to these data as gas in the ``central kpc'' and its corresponding CO intensity $\mathrm{I_{CO}^{NRO}}$ or
$\mathrm{I_{CO}^{IRAM}}$ as $\mathrm{I_{1kpc}}$.  CO
measured in Young et al. (1995) and \cite{KY88} with the 45$^{\prime \prime}$ beamsize
will be referred to as gas in the ``central 3 kpc'', denoted $\mathrm{I_{3kpc}}$.  The
reader should refer to table \ref{tbl-1} for the actual linear scale of the
measurement for individual galaxies.

\clearpage
\clearpage

\begin{figure}
  \begin{center}
  \includegraphics[width=10cm,height=8cm]{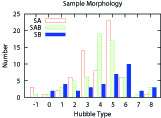}
  \end{center}
  \caption{Morphological distribution of the samples in table \ref{tbl-1} with detected CO in
 the central kpc.} \label{morph_distri}
\end{figure}

Figure \ref{morph_distri} shows the distribution of morphology for all sample galaxies
detected in CO at the central kpc.
The Hubble types are taken from the HyperLEDA database (Paturel et
al. 2003).  Most galaxies lie on Hubble type 4 - 5, corresponding to Sbc -
Sc on de Vaucouleur's scale.  No apparent biases are seen in regard to
the fraction of barred/non-barred galaxies in certain Hubble type,
although we have significantly less SB galaxies compared to SA or SAB
galaxies in total.  A weak bias may exist that more barred galaxies are observed
in the latest types, but may also be representative of the general
galaxy population (e.g., \cite{abraham}).

\clearpage
\clearpage

The inclination ($i$) of the target galaxies result in a variation in the
projected area of the observing beam.  This is especially important in
our case because we observed the galaxies with only one beam.  It is
common to multiply the observed CO intensity by $\cos i$
to correct for this effect. 

\begin{figure}[htbp]
  \begin{center}
  \includegraphics[width=10cm,height=10cm,keepaspectratio]{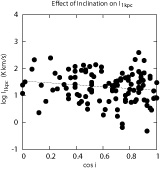}
  \end{center}
  \caption{Variation of CO intensity with inclination $i$ (0 being face-on).  The
 line is a least squares fit with a slope of $-0.35$.  Galaxies those which do
not have normal morphology (peculiars, irregulars, and the like) have
 not been included. } \label{inclination}
\end{figure}

Figure \ref{inclination} shows the variation of CO intensity on $i$.
The observed CO intensity depends very weakly on $i$. If the true CO distribution within the observing 
beam is disk-like, requiring a correction for inclination, the observed CO intensity must increase as the 
inclination becomes edge-on; the effective observing area of the subtended beam becomes larger, scaling as $\propto \cos i^{-1}$.
This will result in a slope of $-1$ in the $\log \mathrm{I_{1kpc}}$ vs. $\cos i$ figure.  A least squares fit to figure \ref{inclination}
results in a slope of $-0.35\pm 0.19$.  Although a variation of CO intensity on inclination does exist, the intrinsic 
dispersion in the CO intensities are far larger.

 This can be explained if the effect is
of geometrical quality, where $i$ in the literature is physically not well-defined
in the central regions, owing to thickness of the disk.  The molecular
gas distribution in the 16$^{\prime \prime}$ beam may not be represented
well by a disk structure, in which case a correction by factor $\cos i$
will be artificial.  The view is also supported by the fact that the
trend in figure \ref{inclination} is same at different wavelengths (\cite{komugi07}), which cannot be explained if the bias
 results from issues with optical depth.

\clearpage
\clearpage

In the succeeding sections,
 we either use the CO intensity alone (without correcting for
 inclination), or use intensity ratios in different beamsizes to cancel
 any effects from inclinations.

\section{Central Concentration and Bars}

\begin{figure}[htbp]
  \begin{center}
  \includegraphics[width=15cm,height=10cm]{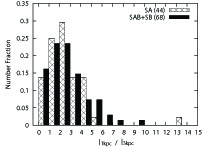}
  \end{center}
  \caption{Number fraction distribution of central concentration of molecular gas for all samples with
 16$^{\prime \prime}$ (or 22$^{\prime \prime}$) and 45$^{\prime \prime}$
 data.  Numbers in the parenthesis are the sample size.  Note the tail of
 high
 concentration for barred galaxies.  One non-barred S0 galaxy, NGC\
 4438, shows $ \mathrm{I_{1kpc}}/\mathrm{I_{3kpc}}=13$, which 
is more than the formal upper limit determined by the area of the two
 observing beams.  This can be attributed to the uncertainties of the observations.} \label{bars_Iratio_all}
\end{figure}

Figure \ref{bars_Iratio_all} shows the number distribution of the ratio
of molecular gas surface mass density in the
central 1 kpc to the central 3 kpc, or the central concentration,
normalized to sample size.  The central concentration of gas in both
barred and non-barred galaxies peak at about 1 to 2, but non-barred
galaxies quickly drop off above 5.  Barred galaxies, however, seem to show a
tail of high central molecular concentration up to 9.  Above 5, barred galaxies dominate
the distribution.  Barred galaxies from \cite{kuno} also show a higher central concentration, and figure
\ref{bars_Iratio_all} confirms this using a large number of samples.  
The probability that the distribution of barred galaxies and that of non-barred galaxies being derived from 
a same parent population (the Kolmogorov-Smirnov (KS) test) is $P_\mathrm{KS} =0.72$.  This indicates that the barred 
samples cannot be said to be significantly different from non-barred
samples.

%  One non-barred S0 galaxy, NGC\ 4438,
% has a central concentration of 13, shown as the most centrally
% concentrated sample in figure 8.  The area subtended 
%by the two beamsizes differ by 4 at most (the higher resolution observation was conducted on the IRAM 30m telescope), giving
% the maximum value indicating that all of the gas is in the smaller beam.  The value probably results from observational errors,
%which can give values outside the formal range.  Furthermore, \cite{KY88} point out that this galaxy has a disturbed bulge.  Since our
%current aim is to characterize the central concentration of gas due to bulges and bar potentials, this galaxy is not sufficient to
%address this point.  Without NGC\ 4438, the probability becomes $P_\mathrm{KS} =0.53$.  However, one outlier galaxy should not affect
%the results of a statistical test severly; if it does, the result is not probably not significant anyways.  We therefore include this
%galaxy in the following series of KS tests.

\begin{figure}[htbp]
  \begin{center}
  \includegraphics[width=15cm,height=10cm]{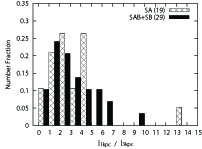}
  \end{center}
  \caption{Same as figure \ref{bars_Iratio_all}, but for a subsample of
 galaxies within the distance range of 13 Mpc to 19 Mpc.  Most are members
 of the Virgo cluster.  Barred galaxies dominate the central
 concentration above 5.} \label{bars_Iratio_virgo}
\end{figure}

 To check for any artificial effects by using samples at
different distances (therefore probing different regions), figure \ref{bars_Iratio_virgo} shows the same figure
but only for samples within the distance range of 13 Mpc to 19 Mpc.  The
trend of barred galaxies having a high central concentration tail is
still seen, but with $P_\mathrm{KS} =0.76$  (see Discussion for why the barred samples may not be observed to have higher central concentration as in previous studies).  
It is interesting that the central concentration in barred galaxies peak
at around the same value as 
non-barred galaxies, but the median of the central concentration is higher for the barred galaxies.
This cannot be explained by the classical picture of bars
simply helping the inflow of molecular gas towards the center, because
then we should expect barred galaxies to have a peak at a higher degree of
central concentration. 

\section{Central Concentration and Hubble Type}

\begin{figure}[htbp]
  \begin{center}
  \includegraphics[width=12cm,height=9cm]{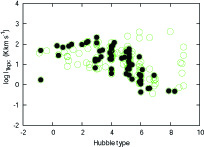}
  \includegraphics[width=12cm,height=9cm]{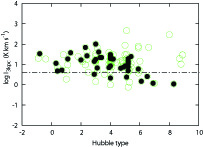}
  \end{center}
  \caption{\footnotesize Top: Distribution of CO intensity in the central 1 kpc with Hubble type.  Filled circles
 are galaxies within the distance of 13 to 19 Mpc, show a somewhat tighter relation which can probably be attributed to
 certainty in distance.  Bottom: Same as top figure, for a the central 3 kpc.  The trend with Hubble type is not
 apparent here.} \label{gas_type}
\end{figure}

Figure \ref{gas_type} shows the CO intensity along the Hubble types, for different beamsizes.  Molecular gas is
significantly more massive in the central 1 kpc of early type galaxies (upper figure), where the large
stellar bulge can be assumed to create a deep potential well.  On the other hand, the
lower figure of figure
\ref{gas_type} shows CO intensity observed with the 45$^{\prime \prime}$ beam, or the central 3 kpc.  It
 is apparent that molecular gas is
not as well correlated with Hubble type than the central 1 kpc.  The typical detection threshold in the FCRAO sample
is $\sim \mathrm{I_{3kpc}} = 2 (\mathrm{K\ km\ s^{-1}})$ ($0.3$ on the logscale), shown in figure \ref{gas_type} with
 the horizontal dashed line.
Many of the CO intensities in the late type galaxies are close to this threshold, but a weak correlation with Hubble
type is still observed.

\begin{figure}[htbp]
  \begin{center}
  \includegraphics[width=15cm,height=10cm]{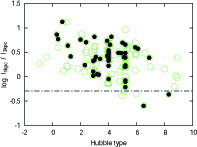}
  \end{center}
  \caption{Ratio of CO intensity within the central 1 kpc and 3 kpc.  Open circles for all samples, and filled
 circles for galaxies within the distance range 13 Mpc to 19 Mpc.  Considering the factor 2-3 difference in
 beamsize (factor 9 difference in subtending beam area), the upper limit
 for the ratio is 9; values above this in the figure can be attributed
 to errors in the measurements, and should not be considered real.} \label{gas_concentration}
\end{figure}

This
indicates that the trend in figure \ref{gas_type}(top) is not created by
the overall low molecular gas mass in the central regions of late type
galaxies, but that the gas distribution is more extended in late type
galaxies.  Figure \ref{gas_concentration} shows the ratio of CO
intensity observed with different beamsizes, and we see that the
distribution of molecular gas is a smooth function of Hubble type, in
that late type galaxies have more extended gas in the center.  For the 
earliest galaxies (Hubble type $\sim 0$, corresponding to S0/S0a), the
intensity ratio $\log \mathrm{I_{1kpc}}/\mathrm{I_{3kpc}}$ is close to 1, where virtually all the
gas is confined to the central 1 kpc.  For the latest types, however,
$\log \mathrm{I_{1kpc}}/\mathrm{I_{3kpc}}$ 
 falls to 0 indicating that there is no apparent central
concentration of gas comparing the central 1 kpc and 3 kpc, so that the
molecular gas disk is likely extended beyond the central 3 kpc.  The horizontal dashed line in figure \ref{gas_concentration}
is the threshold below which non-detections in the FCRAO sample increase, assuming a typical $\log \mathrm{I_{1kpc}} = 0$
for the latest types and $\log \mathrm{I_{3kpc}} = 0.3$ for the FCRAO sample.  In case a bias exists in the 
FCRAO sample that weak latest type galaxies are not observed, the corresponding galaxies will be plotted above this 
line.  The absence of early type galaxies with low central concentration, however, still points to the reality of 
this trend of lower central concetration towards later type galaxies.

\section{Role of Bars within Hubble Types}

The previous sections have revealed that, Hubble type can play an
important role in centrally concentrating molecular gas.  Bars are also
known to centrally concentrate gas, although not found to be
statistically significant based on our current data (see the Discussion
section for possible reasons).

The main setbacks of previous studies is that they have not been able to solve the degeneracy between the effect of Hubble type and bars
because of small sample size.  It is of interest, therefore, which (bars or Hubble type) play the
dominant role in concentrating molecular gas. 

\begin{figure}[htbp]
  \begin{center}
  \includegraphics[width=10cm,height=7cm]{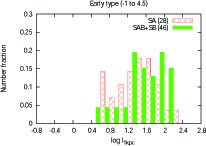}
  \includegraphics[width=10cm,height=7cm]{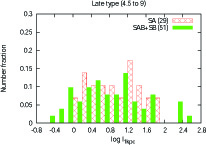}
  \includegraphics[width=10cm,height=7cm]{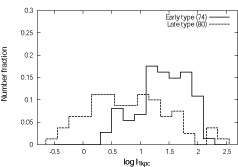}
  \end{center}
  \caption{\footnotesize Distribution of CO intensity for the central 1 kpc for early type (top) and late type (middle) galaxies.
The two groups have a distinctively different distribution.  The bottom figure shows the histogram for the early and late types,
without regard to the presence of bars.} \label{Ico_dist_earlylate}
\end{figure}

Our sample enables the categorization and comparison of different Hubble types and presence of bars,
simultaneously. Figure
\ref{Ico_dist_earlylate}  shows the CO intensity
(molecular gas mass within central 1 kpc)
for early and late type galaxies, respectively, each for barred and
non-barred galaxies.  The bottom figure shows the same figure as the top and the middle, but for barred and non-barred galaxies
combined.  The samples where categorized into early type (type -1 to 4.5) and late type (4.5
to 9) somewhat arbitrary, so that both samples had significant numbers
of galaxies.

If bars concentrate gas into the central region regardless of Hubble type, we can expect $\mathrm{I_{CO}}$ to be higher
in barred galaxies to non-barred galaxies, both in early and late types.  Infact, figure 11 (top) seems to
 suggest that barred early type galaxies
 have a top-heavy molecular mass distribution compared to non-barred early
types with a slightly higher mean CO intensity.  This trend is not statistically siginificant, however, with $P_\mathrm{KS} =0.50$. 
 For barred and non-barred
late type galaxies, $P_\mathrm{KS}$ is  $0.31$.  This is not siginificant either, and it is clear from figure 12 (middle) 
that the apparent difference in the two distribution, results from a wider distribution of CO intensity in barred late type galaxies.
Four barred late type galaxies have
high molecular gas mass ($\log \mathrm{I_{CO}} \ge 2.0$).  All of these
are  known starburst galaxies (NGC\ 253, IC\ 342, NGC\ 3079 and NGC\
6946).  Except for these four, barred galaxies in late types do not
show a high gas density distribution; barred galaxies in
late types apparently do not have the high gas density expected from gas angular momentum 
redistribution.  In cases where the gas is concentrated, they have extremely concentrated
gas and an associated starburst.  This leads us to speculate that these high central concentration late type starbursts,
may either be on the verge of evolving to early type galaxies if the starburst is strong enough (in the context 
of secular evolution), or just using up the molecular fuel to move its place in figure 12 (middle) to the most less massive 
barred late types, which also shows a distinctive tail from non-barred late types.

If the presence of a bulge and its stellar potential drives molecular gas into the central region regardless of whether they have bars,
we can expect early type non-barred galaxies to have higher $\mathrm{I_{1kpc}}$ than late type non-barred galaxies.  Infact, these two
are likely to be significantly different, with $P_\mathrm{KS} =0.06$.  The average CO intensity in early non-barred galaxies 
($\mathrm{I_{1kpc}}=45.1 \ \rm{K\ km\ s^{-1}}$) is more than
 two times larger
than late non-barred galaxies ($\mathrm{I_{1kpc}}=18.9 \ \rm{K\ km\ s^{-1}}$).  The bulge potential seems very effective in 
concentrating molecular gas into the central 1 kpc.
Barred early type galaxies also siginificantly differ from barred late types, with $P_\mathrm{KS} \le 0.001$.  In the same way,
 with the barred and non-barred galaxies
combined, early type galaxies and late type galaxies are, again, significantly different ($P_\mathrm{KS}\le 0.001$).
  From these analysis, it seems that the bulge potential (Hubble type) is the more effective parameter in changing 
$\mathrm{I_{CO}}$ in the central 1 kpc, than bars. 
  The possibility of early type galaxies 
showing more massive molecular gas in the central 1 kpc only because of stronger bars than late types, is rejected.

  It is also important,
however, to normalize the CO intensity in the central 1 kpc with CO data in a larger area to more accurately
 address the central concentration of molecular gas.

\begin{figure}[htbp]
  \begin{center}
  \includegraphics[width=10cm,height=7cm]{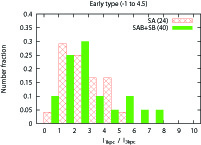}
  \includegraphics[width=10cm,height=7cm]{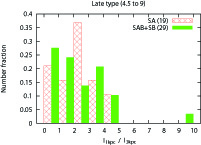}
  \includegraphics[width=10cm,height=7cm]{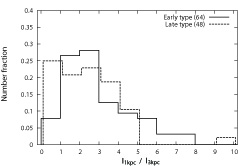}
  \end{center}
  \caption{\footnotesize Distribution of $\mathrm{I_{1kpc}}/\mathrm{I_{3kpc}}$ for early (top)
 and late (middle) type galaxies.  The high central concentration tail seen on the top figure is not seen on the middle.  
The bottom figure shows the histogram for the early and late types, without regard to the presence of bars.} \label{conc_dist_earlylate}
\end{figure}

Figure \ref{conc_dist_earlylate} shows the ratio
of molecular mass in the central 1 kpc to the central 3 kpc, for early and
late type galaxies.  The bottom figure is the same as the top and middle, but with barred and non-barred galaxies combined.
The high central concentration tail (see section 8) in the
distribution can be seen for early type galaxies, but not in late
types (meaning, that early type galaxies constitute the high concentration tail in figure
\ref{bars_Iratio_all}).
The KS test shows that the probability of barred
galaxies and non-barred galaxies for the early type group derived
from the same parent group is $P_\mathrm{KS} =0.62$.  Molecular gas in barred early galaxies are not significantly more centrally
 concentrated than non-barred early types, but the tail of higher concentration in barred galaxies can still be seen, the 
median concentration being 1.6 times higher in barred early types.

The central concentration in late type galaxies in figure 12 (middle) seem much more insensitive to the presence of 
bars.  The tail of higher concentration that was seen in early types are not seen.  
The only exception of a highly centrally
concentrated barred late type galaxy is NGC\ 3486, with practically all of the gas
within its central 1 kpc.  This SABc galaxy is only weakly barred, with a
bar strength (ratio of maximum tangential force and average radial
force) of 0.2 (\cite{laurikainen}).   The central concentration
of molecular gas in this galaxy is extraordinary in terms of bar and
Hubble type, within the context of this paper.
 The KS-test
for barred and non-barred galaxies in the late type group show that the
two morphologies can be
derived from a same distribution by $P_\mathrm{KS} =0.71$ probability.

Comparing non-barred early types and non-barred late types give $P_\mathrm{KS} =0.26$.  Similarly, barred
 early types and barred late types
give $P_\mathrm{KS} =0.39$. 
% Again, these are not significant, but are smaller than the $P_\mathrm{KS}$ when comparing barred and non-barred galaxies
%in a certain type.  This points to the likelihood of Hubble type contributing more to central concentration than bars.

  From the bottom figure of \ref{conc_dist_earlylate}, we see that almost all early type galaxies
have a ratio 
$\mathrm{I_{1kpc}}/\mathrm{I_{3kpc}}$ of more than 1, indicating that the central 1 kpc is almost always more dense than its surrounding.  Late type galaxies,
on the other hand, actually peak at less than 1, meaning that many late type galaxies are less dense in the central kpc compared
to its surrounding disk.  Early type galaxies are more likely to be centrally concentrated than late types ($P_\mathrm{KS} =0.25$).

\section{Discussion}

Central concentration of molecular gas from bars in early type (Sbc and earlier) galaxies had
already been found by Sheth et al. (2005), where early barred galaxies
showed concentration enhancement of a factor 4 compared to early
non-barred galaxies.  The median central concentration for our sample of
early type bars ($\sim$4.1) is also higher than early type non-bars ($\sim$2.6), although not by
a factor of 2.  Our definition of central
concentration $\mathrm{I_{1kpc}}/\mathrm{I_{3kpc}}$ employs the molecular mass
in the central 45$^{\prime \prime}$ for normalization whereas the \cite{sheth} sample uses the global molecular mass.  Since the central
45$^{\prime \prime}$ ($\sim 3$ kpc) would likely not contain all of the galaxy's molecular gas, any trends in the central concentration
would be mitigated compared with normalizations using the whole molecular content.  
  Kuno et al. (2007), similarly, used CO maps to define the central concentration of molecular gas in a sample of 40 galaxies.  They found
barred galaxies are significantly ($P_\mathrm{KS} =0.004$) more centrally concentrated than non-barred galaxies.  Therefore, our result
 that barred galaxies are not significantly more concentrated ($P_\mathrm{KS} =0.76$), is likely to be a result of using the central
 45$^{\prime \prime}$ for normalization rather than the central concentration being not as strong.  Errors in the measurements, which
can change the derived CO intensity by factor two when comparing different telescopes, can also reduce the power of comparison with the
FCRAO measurements.  
These sources of error apply to the whole sample, however, and the observed result that Hubble type affects the central concentration
more effectively than bars, will likely not be changed.

\clearpage


\clearpage

\section{Summary and Conclusions}

This paper outlined a CO survey of nearby galactic centers conducted
with the NRO 45m telescope.  Combined with published data at similar
 and lower resolution, we are able
to infer on the molecular gas distribution in barred and unbarred galaxies
of various Hubble types.  The number of galaxies used ($\sim 160$) is
far more than that used in previous interferometric and single dish
mapping studies.  The main results of this study are; 

\begin{enumerate}
\item
Molecular gas is denser in non-barred early type galaxies compared to non-barred late type galaxies.
Therefore, possible stronger bars in early types cannot be the dominant reason for massive molecular gas observed in early types.
\item
Central concentration of molecular gas in early type galaxies is
     confirmed.  The ratio of molecular
     gas in the central kiloparsec to the central 3 kiloparsecs declines
     smoothly with Hubble type.
\item
Barred early type galaxies have a higher median of central concentration than non-barred early type galaxies,
suggesting a tail distribution of high central concentration in barred
     galaxies, although not statistically significant based on the
     current data.  This trend is not observed in late type galaxies.
\end{enumerate}

%\section{Acknowledgment}
\acknowledgements
The authors thank the anonymous referee for valuable comments and careful reading of the manuscript.
The authors also wish to thank the NRO staff for generous allocation of telescope time.
S.K., S.O., F.E., and K.M. were financially supported by a Research Fellowship
from the Japan Society for the Promotion of Science for Young Scientists.
We acknowledge the usage of the HyperLeda database (http://leda.univ-lyon1.fr), and the NASA/IPAC Extragalactic Database(NED),
which is operated by the Jet Propulsion Laboratory, Caltech, under
contract with the National Aeronautics and Space Administration.
%\clearpage

\end{document}